\documentclass{ametsocV6.1}
\usepackage{amsmath,amsfonts,amssymb,bm}
\usepackage{mathptmx}
\usepackage{newtxtext}
\usepackage{newtxmath}
\usepackage{color}

\newcommand{\yo}[1]{{\color{black} #1}}

\title{Breaking of internal waves parametrically excited by ageostrophic anticyclonic instability}

\authors{Yohei Onuki,\aff{a}\aff{b} \correspondingauthor{Yohei Onuki, onuki@riam.kyushu-u.ac.jp}
Sylvain Joubaud,\aff{b}\aff{c}
Thierry Dauxois\aff{b}}

\affiliation{\aff{a}{Research Institute for Applied Mechanics, Kyushu University, Kasuga, Fukuoka, Japan}\\
\aff{b}{ENS de Lyon, CNRS, Laboratoire de Physique, Lyon, France}\\
\aff{c}{Institut Universitaire de France (IUF), Paris, France}}

\abstract{
A gradient-wind balanced flow with an elliptic streamline parametrically excites internal inertia-gravity waves through ageostrophic anticyclonic instability (AAI). This study numerically investigates the breaking of internal waves and the following turbulence generation resulting from the AAI. In our simulation, we periodically distort the calculation domain following the streamlines of an elliptic vortex and integrate the equations of motion using a Fourier spectral method. This technique enables us to exclude the overall structure of the large-scale vortex from the computation and concentrate on resolving the small-scale waves and turbulence.
From a series of experiments, we identify two different scenarios of wave breaking conditioned on the magnitude of the instability growth rate scaled by the buoyancy frequency, $\lambda/N$. First, when $\lambda/N\gtrsim0.008$, the primary wave amplitude excited by AAI quickly goes far beyond the overturning threshold and directly breaks. The resulting state is thus strongly nonlinear turbulence. Second, if $\lambda/N\lesssim0.008$, weak wave-wave interactions begin to redistribute energy across frequency space before the primary wave reaches a breaking limit. Then, after a sufficiently long time, the system approaches a Garrett-Munk-like stationary spectrum, in which wave breaking occurs at finer vertical scales.
Throughout the experimental conditions, the growth and decay time scales of the primary wave energy are well correlated. However, since the primary wave amplitude reaches a prescribed limit in one scenario but not in the other, the energy dissipation rates exhibit two types of scaling properties. This scaling classification has similarities and differences with D'Asaro and Lien's (2000) wave-turbulence transition model. (248 words)
}

\begin{document}

\maketitle

%
%
%
\statement
Due to the gradients in buoyancy and pressure, density-stratified seawater supports oscillatory vertical motion called internal waves. When waves significantly skew a density isosurface, dense water lifts over lighter water resulting in gravitational instability and high energy dissipation. In this wave-breaking process, seawater is vertically mixed, transporting heat and nutrients essential to maintain the Earth's climate and ecosystems. This study investigates the generation and breaking of ocean internal waves in a novel numerical simulation setup; we temporally distort the model shape to emulate the wave excitation forced by a larger-size horizontal eddy, a \yo{ubiquitous} situation at $O(1$-$10)$km scales in the upper ocean. The simulation results exhibit two unique wave-breaking scenarios with distinct scaling features in turbulence energy dissipation rates. (119 words)
%
%

%
\section{Introduction}
\yo{Motions of stably stratified and rotating fluid are largely classified into two categories; a slowly evolving horizontal flow and a rapidly oscillating three-dimensional flow \citep{yasuda2015theoretical1,yasuda2015theoretical2,chouksey2018internal}. In the former part, the pressure gradient is almost equilibrated with the gravity and Coriolis or centrifugal force, which is why this part is occasionally called the balanced mode (BM). The latter, imbalanced mode identified as the residual of the total flow subtracted by the BM contains internal inertia-gravity waves\textemdash internal gravity waves affected by Coriolis force\textemdash that we simply call inertia-gravity waves (IGWs).}

Although the large-scale circulation of the atmosphere and ocean are dominated by the BM, IGWs play some auxiliary but important roles. A representative one is to enhance the dissipation of energy. In the ocean, kinetic energy originally injected by winds or tides are eliminated by molecular viscosity at the Kolmogorov scale, which is no more than 1 cm length \citep{thorpe2005turbulent}. The scales at which the BMs dominates the flow energy are, on the other hand, larger than tens of kilometers. In addition, energy of the BM tends to be transferred towards \yo{larger} scales through the inverse cascade \citep{scott2005direct}. Therefore, the small-scale turbulence energy apart from the boundary layers may not be supplied directly from the BM but from IGWs. Indeed, observational studies have revealed that the energy spectrum with the horizontal scale less than tens of kilometers is dominated by IGWs \citep{lien2019small}. The transition from the large-scale BM to the smaller-scale IGWs has also been detected in the global-scale ocean models \citep{qiu2018seasonality,torres2018partitioning}.

The rate of energy transfer between the BM and IGWs depends on the Rossby number, $Ro = \mathcal{U} / (f\mathcal{L})$, where $f$ is the Coriolis parameter and $\mathcal{U}$ and $\mathcal{L}$ are the typical velocity and a length scale of the fluid motion. When $Ro \ll 1$, the BM and IGWs behave almost independently. When $Ro$ is comparable to or greater than unity, nonlinear coupling between the BM and IGWs takes place. In the ocean, $Ro$ increases as a spatial scale is reduced and reaches unity at a horizontal length of $O(1) \sim O(10)$ km named the submesoscales \citep{mcwilliams2016submesoscale}. Interaction between the BM and IGWs at the submesoscales crucially influences the ocean energy budget yet remains poorly quantified.

A difficulty in understanding submesoscale wave-flow coupling is the presence of external forces. For example, heating/cooling and precipitation/evaporation at the ocean surface change the density of seawater, which induces convective motions in the mixed layer. Horizontal density gradients in the mixed layer are further intensified by the \yo{horizontal velocity shear}, creating a sharp density front and the associated thermal-wind-balanced flow \citep{mcwilliams2009cold}. Wind forcing at the sea surface drives near-inertial waves in the mixed layer, which subsequently exchange energy with the ambient frontal motion \citep{thomas2017modifications}. Many high-resolution numerical experiments have demonstrated complication of these entangled processes \citep[e.g.,][]{mahadevan2006analysis,capet2008mesoscale1,capet2008mesoscale2,capet2008mesoscale3,molemaker2005baroclinic,barkan2017stimulated}.

To gain a better insight into the highly complicated submesoscale dynamics, isolating a particular phenomenon excluding external forces and boundary effects is effective. Following this direction, this study investigates a process named ageostrophic anticyclonic instability (AAI) in an idealized setting. In AAI, IGW disturbances are exponentially amplified from a fully balanced anticyclonic flow field. AAI looks similar to other submesoscale instabilities, specifically symmetric instability (SI) and centrifugal instability (CI), but they are distinguished in terms of the instability criterion. In order for CI or SI to occur, the sign of the Ertel's potential vorticity must be reversed. Since the potential vorticity is adiabatically conserved along a stream, this condition is met only when a strong external force or a viscous drag at a boundary layer is present \citep[e.g.,][]{thomas2010reduction,gula2016topographic}. For AAI, there is no such a clear threshold; i.e., almost any kind of BM can generate IGWs, but its intensity sharply depends on the Rossby number. An asymptotic analysis proves that the amplitude of the IGWs generated from a geostrophically balanced flow involves a factor ${\rm e}^{- C/ Ro}$, where $C$ is some constant independent of $Ro$ \citep{vanneste2004exponentially,vanneste2008exponential,vanneste2013balance}. This property makes AAI insignificant at the mesoscales or larger where $Ro$ is small. At submesoscales where $Ro \sim O(1)$, it is expected that AAI ubiquitously occurs to assist energy transfer from the BM to IGWs.

The onset of AAI is controlled not only by $Ro$ but also by the flow geometry. When one writes the absolute vorticity and the strain rate associated with the BM as $A$ and $S$, respectively, AAI becomes significant when $A - S \lesssim 0$. In light of this condition, AAI has been analyzed in a variety of situations. \cite{mcwilliams1998fluctuation} reported that a barotropic anticyclonic vortex with its streamline an elliptic shape parametrically excites \yo{IGWs}. \cite{mcwilliams2004ageostrophic} found that ageostrophic instability arises at a coastal boundary current via resonant interaction between an IGW and a coastal Kelvin wave. \cite{molemaker2005baroclinic} and \cite{wang2014ageostrophic} investigated ageostrophic instabilities of geostrophically balanced vertically sheared flow and compared them with the usual geostrophic baroclinic instability. \cite{menesguen2012ageostrophic} explored ageostrophic instability of a localized interior jet. Among these, this study considers the first case, instability of an elliptic vortex, the simplest configuration that may commonly apply to a wide area of the world ocean.

Classically, it has been known that three-dimensional oscillating disturbances are amplified within a two-dimensional elliptic vortex \citep{kerswell2002elliptical}. This type of instability, particularly named the elliptic instability, is a ubiquitous process in turbulent energy cascade \citep{mckeown2020turbulence}. Although a stability analysis of a finite-size elliptic vortex is possible, an assumption that the size of the vortex is very large and the velocity gradient is locally homogeneous makes the problem much simpler. In that case, the local stability analysis approach can be employed \citep[e.g.,][]{bayly1986three,lifschitz1991local,waleffe1990three,landman1987three,craik1986evolution,craik1989stability,craik1992stability,leblanc1997stability,ghaemsaidi2019three}. In this approach, a sinusoidal ansatz whose wavenumber rapidly changes is assumed. The temporal evolution of the amplitude and wavenumber of this ansatz is described by a small set of ordinary differential equations. Once the wavenumber is obtained as a function of time, the stability of the amplitude equation is analyzed based on Floquet theorem. Using this method, \cite{miyazaki1993elliptical}, \cite{miyazaki1992three}, and \cite{mcwilliams1998fluctuation} investigated the stability of an elliptic vortex in a stably stratified fluid with and without rotation. Later, \cite{aspden2009elliptical} presented a rigorous asymptotic consideration for the small $Ro$ regime and found the aforementioned exponentially small instability growth rate.

To the authors' knowledge, studies of elliptic instability or AAI arising in the ocean or atmosphere are scarce and mostly limited to linear stability analysis. Linear theory is inherently incapable of addressing turbulent dissipation resulting from instability. Furthermore, the instability growth rate derived from linear analysis is not necessarily a good indicator even for the basic energy supply to unstable disturbances \citep{onuki2018decay}. Nonlinear analysis is thus essential to reveal the roles of AAI on ocean energetics.

In this study, we explore long-term behaviors of IGWs parametrically excited by AAI within an elliptic vortex using fully nonlinear numerical simulations. To resolve the breaking of IGWs and the following turbulence generation most precisely while incorporating energy supply from a much larger-scale vortex, we utilize the domain distortion technique combined with Fourier-spectral discretization. This technique was applied for rotating or stratified fluid by \cite{le2017inertial} and \cite{le2018parametric}. Since the scope of these past studies was the parametric wave excitation due to tidal deformation in planetary cores, the distortion was so small that the resulting state is weak turbulence in which waves do not break. This study, on the other hand, deals with the breaking processes of internal gravity waves that occur in the oceanic conditions. The methodology is common with that of the work recently presented by the authors, \cite{onuki2021simulating}, that investigated breaking of small-scale internal waves parametrically excited by a larger-scale internal wave.

The plan of this paper is as follows. The model formulations and some fundamentals of the elliptic instability are described in section 2. We carry out fully nonlinear numerical simulations and demonstrate the wave breaking mechanisms in Section 3. The parameter dependence of the energy budgets are discussion in section 4. Section 5 presents the conclusions.

\section{The model equation and its linear stability}
We consider a density stratified fluid in a rotating frame, governed by a set of equations,
\begin{subequations} \label{eq:equation_boussinesq}
\begin{align}
\partial_t \boldsymbol{u} + \boldsymbol{u} \cdot \nabla \boldsymbol{u} + f \boldsymbol{e}_z \times \boldsymbol{u} & = \frac{- \nabla p - g \rho \boldsymbol{e}_z}{\rho_{\rm ref}} + \nu_u \nabla^2 \boldsymbol{u} \label{eq:equation_of_motion} \\
\partial_t \rho + \boldsymbol{u} \cdot \nabla \rho & = \nu_\rho \nabla^2 \rho \\
\nabla \cdot \boldsymbol{u} & = 0 , \label{eq:incompressible}
\end{align}
\end{subequations}
where $\boldsymbol{u} = (u, v, w)$ is the velocity vector, $\rho$ is the density, $p$ is the pressure, $\rho_{\rm ref}$ is a constant reference density, $f$ is the Coriolis parameter assumed to be a positive constant, $\nu_u$ and $\nu_\rho$ are kinematic viscosity and diffusivity, respectively, and $\boldsymbol{e}_z$ is the upward unit vector. The spatial coordinates are defined as $\boldsymbol{x} = (x, y, z)$ with $x$ and $y$ specifying the horizontal directions and $z$ the vertical direction. We have used the Boussinesq approximation to exclude the variations in inertia originating from density difference.

The interest of this study is directed to the behavior of small-scale disturbances within a vertically homogenous balanced motion. The scale-separation assumption and a proper coordinate change simplify such a reference state to be represented by a linear stratification, namely a constant Brunt-V\"ais\"al\"a frequency $N$, and a quadratic form of streamfunction, a common model with that of \cite{mcwilliams1998fluctuation}, as
\begin{align}
\psi = - \frac{\alpha x^2 + \beta y^2}{2}, \quad \beta \geq \alpha .
\end{align}
It is verified that
\begin{align}
\boldsymbol{u} = \boldsymbol{U}(x, y) = (U, V, 0), \quad \rho = \overline{\rho}(z), \quad \mbox{and} \quad p = P(x,y,z),
\end{align}
with $(U, V) \equiv (-\partial_y \psi, \partial_x \psi) = (\beta y, - \alpha x)$, $\overline{\rho} \equiv \rho_{\rm ref} \left( 1 - N^2 z / g \right)$ and $P \equiv \rho_{\rm ref} \left[ - \alpha (f - \beta)x^2 /2 - \beta (f - \alpha)y^2 / 2 - \yo{(gz - N^2 z^2 / 2)} \right]$, compose a gradient-wind balanced stationary solution of the original set of governing equations (\ref{eq:equation_boussinesq}). In order for the streamline to close, $\alpha$ and $\beta$ should have a same sign. Then, we suppose that $\beta \geq \alpha > 0$, which corresponds to a zonally elongated anticyclonic vortex. Although the following discussion also applies to a cyclonic vortex, $\alpha, \beta < 0$, the instability is much weaker in that case.

When one considers a finite-size vortex, the relative vorticity well represents the typical velocity, $\mathcal{U}$, divided by a characteristic length scale, $\mathcal{L}$. From this reason, we identify the Rossby number in the present model as the absolute value of the relative vorticity in a reference state, $\partial_x V - \partial_y U = - \alpha - \beta$, divided by the planetary vorticity \citep{barkan2017stimulated}, i.e.,
\begin{align}
Ro \equiv \frac{\alpha + \beta}{f} .
\end{align}
If $Ro > 1$, the sign of the absolute vorticity is opposite to that of the planetary vorticity. To exclude the centrifugal instability from the consideration, we further impose a restriction of $Ro \leq 1$. Next, we specify the geometrical character of the flow field using the ellipticity,
\begin{align} \label{eq:ellipticity}
e = 1 - \sqrt{\frac{\alpha}{\beta}} .
\end{align}
The range of this parameter is $0 \leq e < 1$. A circular vortex corresponds to $e=0$. In the limit of $e \to 1$, the stream geometry approaches to a parallel shear flow.

To investigate the stability of the system, we superimpose three-dimensional disturbances on the reference stream. Inserting $\boldsymbol{u} = \boldsymbol{U} + \boldsymbol{u}' = (U + u', V + v', w'), \rho = \overline{\rho} - \rho_{\rm ref} N \theta' / g, p = P + \rho_{\rm ref} p'$ into (\ref{eq:equation_boussinesq}), we derive the governing equations for the disturbance components as
\begin{subequations} \label{eq:equation_disturbances}
\begin{align}
D_t \boldsymbol{u}' + \boldsymbol{u}' \cdot \nabla\boldsymbol{U} + \boldsymbol{u}' \cdot \nabla \boldsymbol{u}' + f \boldsymbol{e}_z \times \boldsymbol{u}' & = - \nabla p' + N \theta' \boldsymbol{e}_z + \nu_u \nabla^2 \boldsymbol{u}' \label{eq:disturbance_u} \\
D_t \theta' + \boldsymbol{u}' \cdot \nabla \theta' + N w' & = \nu_\rho \nabla^2 \theta' \label{eq:disturbance_theta} \\
\nabla \cdot \boldsymbol{u}' & = 0 ,  \label{eq:disturbance_incompressible}
\end{align}
\end{subequations}
where $D_t \equiv \partial_t + \boldsymbol{U} \cdot \nabla$ represents the temporal differentiation along the reference stream, the perturbation pressure $p'$ is scaled by $\rho_{\rm ref}$ to make the expression concise, and $\theta'$ is the buoyancy perturbation of the velocity unit. We regard the set of equations, (\ref{eq:equation_disturbances}), as the basic model of the study and, in the remaining of this section, analyze its energetics, kinematic character and linear stability.

The kinetic energy and the available potential energy of the disturbance components integrated over the whole volume are now $E_u \equiv (1/2) \int |\boldsymbol{u}'|^2 d\bm{x}$ and $E_\theta \equiv (1/2) \int \theta'^2 d\bm{x}$, respectively. Neglecting energy flux passing through the boundary, we describe the budgets of energy by
\begin{subequations} \label{eq:budgets}
\begin{align}
\frac{d E_u}{dt} & = \mathcal{P} - \mathcal{C} - \epsilon_u \\
\frac{d E_\theta}{dt} & = \mathcal{C} - \epsilon_\theta ,
\end{align}
\end{subequations}
where, $\mathcal{P}$ is the lateral shear production rate, $\mathcal{C}$ is the energy conversion rate from the kinetic energy to available potential energy, and $\epsilon_u$ and $\epsilon_\theta$ are the dissipation rates of kinetic energy and available potential energy, respectively. They are specifically represented as
\begin{subequations}
\begin{align}
\mathcal{P} & \equiv \int (\alpha - \beta)u'v' d\bm{x} \label{eq:energy_production} \\
\mathcal{C} & \equiv - \int N w' \theta' d\bm{x} \\
\epsilon_u & \equiv \int \nu_u \left( |\nabla u'|^2 + |\nabla v'|^2 + |\nabla w'|^2 \right) d\bm{x} \\
\epsilon_\theta & \equiv \int \nu_\rho |\nabla \theta'|^2 d\bm{x} .
\end{align}
\end{subequations}
Since the budget of the total energy is $d(E_u + E_{\yo{\theta}})/dt = \mathcal{P} - \epsilon_u - \epsilon_\theta$, and $\epsilon_u$ and $\epsilon_\theta$ are both positive, energy injected into the system via $\mathcal{P}$ should be compensated by the dissipation through $\epsilon_u$ and $\epsilon_\theta$ in a stationary state.

\subsection{Fourier analysis}
Now, for an initial condition of the variables, $(\boldsymbol{u}', \theta', p')$, let us decompose it into a Fourier integral with respect to the wavenumber, $\tilde{\boldsymbol{k}} = (\tilde{k}, \tilde{\ell}, \tilde{m})$, as
\begin{align}
(\boldsymbol{u}'(\boldsymbol{x}, 0), \theta'(\boldsymbol{x}, 0), p'(\boldsymbol{x}, 0)) = \frac{1}{(2\pi)^{3/2}} \int (\hat{\boldsymbol{u}}(\tilde{\boldsymbol{k}}, 0), \hat{\theta}(\tilde{\boldsymbol{k}}, 0), \hat{p}(\tilde{\boldsymbol{k}}, 0)) {\rm e}^{{\rm i} \tilde{\boldsymbol{k}} \cdot \boldsymbol{x}} d \tilde{\boldsymbol{k}} .
\end{align}
We then consider the evolutions of the Fourier components, $(\hat{\boldsymbol{u}}, \hat{\theta}, \hat{p})$. At this stage, a difference from the usual computation of the Fourier spectral method arises. Since the differential operator $D_t$ contains spatially inhomogeneous factors, each Fourier component is not linearly independent. To solve this problem, we introduce a time-dependent wavenumber, $\boldsymbol{k}(t) = (k, \ell, m)$, and postulate the following form of expression,
\begin{align}
D_t (\hat{\boldsymbol{u}} {\rm e}^{{\rm i} \boldsymbol{k} \cdot \boldsymbol{x}}) = (\partial_t \hat{\boldsymbol{u}}) {\rm e}^{{\rm i} \boldsymbol{k} \cdot \boldsymbol{x}} .
\end{align}
In order for this expression to be identically valid, the time-dependent wavenumbers should satisfy
\begin{align} \label{eq:equation_k}
\frac{d \boldsymbol{k}}{d t} = - \nabla \boldsymbol{U} \cdot \boldsymbol{k},
\end{align}
i.e., $dk / dt = \alpha \ell$, $d\ell / dt = - \beta k$ and $dm / dt = 0$. Solving these equations, the wavenumbers are determined as
\begin{align} \label{eq:variation_in_k}
k = \tilde{k} \cos \omega_v t + \frac{\tilde{\ell}}{r} \sin \omega_v t, \quad
\ell = - r \tilde{k} \sin \omega_v t + \tilde{\ell} \cos \omega_v t, \quad
m = \tilde{m} ,
\end{align}
where $\omega_v \equiv \sqrt{\alpha \beta}$ and $r \equiv \sqrt{\beta / \alpha}$ represent the rotation frequency and the long/short axes ratio of the vortex, respectively. To make the dependence of the temporally varying wavenumber $\boldsymbol{k}$ on the initial wavenumber $\tilde{\boldsymbol{k}}$ explicit, it would be better to write it as $\boldsymbol{k} = \boldsymbol{k}(\tilde{\boldsymbol{k}}, t)$. Consequently, Fourier transform at an arbitrary time becomes
\begin{align} \label{eq:Fourier_time_dependent}
(\boldsymbol{u}'(\boldsymbol{x}, t), \theta'(\boldsymbol{x}, t), p'(\boldsymbol{x}, t)) = \frac{1}{(2\pi)^{3/2}} \int (\hat{\boldsymbol{u}}(\tilde{\boldsymbol{k}}, t), \hat{\theta}(\tilde{\boldsymbol{k}}, t), \hat{p}(\tilde{\boldsymbol{k}}, t)) {\rm e}^{{\rm i} \boldsymbol{k}(\tilde{\boldsymbol{k}}, t) \cdot \boldsymbol{x}} d \tilde{\boldsymbol{k}} .
\end{align}
The governing equations in Fourier space are
\begin{subequations} \label{eq:governing_equations_k}
\begin{align}
\partial_t \hat{\boldsymbol{u}} + \hat{\boldsymbol{u}} \cdot \nabla \boldsymbol{U} + \mathcal{F}[\boldsymbol{u}' \cdot \nabla \boldsymbol{u}'] + f \hat{\boldsymbol{e}}_z \times \hat{\boldsymbol{u}} & = - i \boldsymbol{k} \hat{p} + N \hat{\theta} \hat{\boldsymbol{e}}_z - \nu_u |\boldsymbol{k}|^2 \hat{\boldsymbol{u}} \label{eq:disturbance_k_u} \\
\partial_t \hat{\theta} + \mathcal{F}[\boldsymbol{u}' \cdot \nabla \theta'] + N \hat{w} & = - \nu_\rho |\boldsymbol{k}|^2 \hat{\theta} \label{eq:disturbance_k_theta} \\
\boldsymbol{k} \cdot \hat{\boldsymbol{u}} & = 0 ,  \label{eq:disturbance_k_incompressible}
\end{align}
\end{subequations}
where nonlinear terms are represented using the conventional notation of Fourier transform, $\mathcal{F}[ \cdot ] \equiv 1 / (2\pi)^{3/2} \int \cdot {\rm e}^{-{\rm i} \boldsymbol{k}(\tilde{\boldsymbol{k}}, t) \cdot \boldsymbol{x}} d \boldsymbol{x}$. Although in these equations the independent coordinates are chosen as $(\tilde{\boldsymbol{k}}, t)$, in the energetic analysis, it would be better to use the {\it genuine} wavenumber $\boldsymbol{k}$ instead of the {\it initial} wavenumber $\tilde{\boldsymbol{k}}$. For this purpose, we also write the functional relationship between $\boldsymbol{k}$ and $\tilde{\boldsymbol{k}}$ as $\tilde{\boldsymbol{k}} = \tilde{\boldsymbol{k}}(\boldsymbol{k}, t)$ and define the kinetic and available potential energy spectra as
\begin{align}
\hat{E}_u(\boldsymbol{k}, t) = \frac{|\hat{\boldsymbol{u}}(\tilde{\boldsymbol{k}}(\boldsymbol{k}, t), t)|^2}{2} \quad \mbox{and} \quad \hat{E}_\theta(\boldsymbol{k}, t) = \frac{|\hat{\theta}(\tilde{\boldsymbol{k}}(\boldsymbol{k}, t), t)|^2}{2},
\end{align}
respectively. Then, spectral energy budgets are derived as
\begin{subequations} \label{eq:budgets_wavenumber}
\begin{align}
\frac{\partial \hat{E}_u}{\partial t} + \nabla_k \cdot (\dot{\boldsymbol{k}} \hat{E}_u) & = \underbrace{- \Re[\hat{\boldsymbol{u}}^\dag \cdot (\hat{\boldsymbol{u}} \cdot \nabla) \boldsymbol{U}]}_{\hat{\mathcal{P}}} + \underbrace{\Re[N \hat{w}^\dag \hat{\theta}]}_{-\hat{\mathcal{C}}} \underbrace{- \Re[\hat{\boldsymbol{u}}^\dag \cdot \mathcal{F} [(\boldsymbol{u}' \cdot \nabla) \boldsymbol{u}']]}_{\hat{\mathcal{T}}_u} - \underbrace{2 \nu_u |\boldsymbol{k}|^2 \hat{E}_u}_{\hat{\epsilon}_u} \\
\frac{\partial \hat{E}_\theta}{\partial t} + \nabla_k \cdot (\dot{\boldsymbol{k}} \hat{E}_\theta) & =  \underbrace{- \Re[N \hat{w}^\dag \hat{\theta}]}_{\hat{\mathcal{C}}} \underbrace{- \Re[\hat{\theta}^\dag \mathcal{F} [(\boldsymbol{u}' \cdot \nabla) \theta']]}_{\hat{\mathcal{T}}_\theta} - \underbrace{2 \nu_\rho |\boldsymbol{k}|^2 \hat{E}_\theta}_{\hat{\epsilon}_\theta} ,
\end{align}
\end{subequations}
where $\nabla_k$ is the gradient operator in wavenumber space, $\dot{\boldsymbol{k}} \equiv d\boldsymbol{k} / dt$ is the temporal differentiation of a wavevector, ${}^\dag$ represents the complex conjugate, and $\Re$ denotes taking the real part. In each expression, a linear energy flux divergence term appearing on the left-hand side reflects the variations in wavenumber induced by the reference stream as specified by (\ref{eq:variation_in_k}). On the right-hand sides, $\hat{\mathcal{P}}$ is the lateral shear production rate, $\hat{\mathcal{C}}$ is the conversion rate from the kinetic to available potential energy, $\hat{\mathcal{T}}_u$ and $\hat{\mathcal{T}}_\theta$ are the nonlinear energy transfer rates, and $\hat{\epsilon}_u$ and $\hat{\epsilon}_\theta$ are the energy dissipation rates, respectively. Finally, integrations of (\ref{eq:budgets_wavenumber}) over wavenumber space coincide with the total energy budgets, (\ref{eq:budgets}), as the termwise correspondence,
\begin{align}
\int (\hat{E}_u, \hat{E}_\theta, \hat{\mathcal{P}}, \hat{\mathcal{C}}, \hat{\mathcal{T}}_u, \hat{\mathcal{T}}_\theta, \hat{\epsilon}_u, \hat{\epsilon}_\theta) d\boldsymbol{k} = (E_u, E_\theta, \mathcal{P}, \mathcal{C}, 0, 0, \epsilon_u, \epsilon_\theta) ,
\end{align}
is established.

\yo{From now on until the end of this section}, we assume that the disturbance amplitudes are sufficiently small such that the nonlinear terms are negligible. This postulate makes each wavenumber component completely independent. In addition, we shall set the inviscid condition, $\nu_u = \nu_\rho = 0$, for the moment. Consequently, the system is governed by the following set of ordinary differential equations,
\begin{subequations} \label{eq:linear}
\begin{align}
\frac{d\hat{u}}{dt} & = (f - \beta) \hat{v} - {\rm i} k \hat{p}  \label{eq:lu} \\
\frac{d\hat{v}}{dt} & = (\alpha - f) \hat{u} - {\rm i} \ell \hat{p}  \label{eq:lv} \\
\frac{d\hat{w}}{dt} & = - {\rm i} m \hat{p} + N \hat{\theta}  \label{eq:lw} \\
\frac{d\hat{\theta}}{dt} & =  - N \hat{w}  \label{eq:lt} \\
0 & = k \hat{u} + \ell \hat{v} + m \hat{w}.  \label{eq:incompressible_l}
\end{align}
\end{subequations}
The state of the system is now specified by four complex variables, $\hat{u}, \hat{v}, \hat{w}$ and $\hat{\theta}$. Taking into account the incompressibility condition (\ref{eq:incompressible_l}), the number of degrees of freedom is three. Note that multiplying (\ref{eq:lu}), (\ref{eq:lv}), and (\ref{eq:lw}) by $k$, \yo{$\ell$}, and $m$, respectively, and using (\ref{eq:incompressible_l}), one may obtain $\hat{p}$ at any instance as a function of other variables.

\subsection{Wave-vortex decomposition}
If the background stream is absent, $\alpha = \beta = 0$, it is known that the motion is decomposed into a pair of oscillating IGWs and one stationary BM, which are normally referred to as the wave and vortical modes, respectively \citep{lien1992normal,herbert2016waves}. Here, we extend this wave-vortex decomposition technique into a situation of finite $\alpha$ and $\beta$.

Now, (\ref{eq:linear}) leads to the potential vorticity conservation, $d\hat{q} / dt = 0$, with
\begin{align} \label{eq:q_lin}
\hat{q}(\hat{\boldsymbol{u}}, \hat{\theta}) = {\rm i} N (k \hat{v} - \ell \hat{u}) + {\rm i} \Omega m \hat{\theta},
\end{align}
where $\Omega \equiv f - \alpha - \beta$ is the absolute vorticity of the reference state. To decompose the flow field into the wave and the vortical modes uniquely, $(\hat{\boldsymbol{u}}, \hat{\theta}) = (\hat{\boldsymbol{u}}_w, \hat{\theta}_w) + (\hat{\boldsymbol{u}}_v, \hat{\theta}_v)$, we raise the following postulates: (i) the vortical mode components, $(\hat{\boldsymbol{u}}_v, \hat{\theta}_v)$, are proportional to $\hat{q}$, and (ii) the wave mode and vortical mode are energetically orthogonal. To begin with, (i) yields that the vortical modes are formally expressed as
\begin{align}
(\hat{u}_v, \hat{v}_v, \hat{w}_v, \hat{\theta}_v) = (\gamma_u, \gamma_v, \gamma_w, \gamma_\theta) \hat{q} ,
\end{align}
and we would like to know the coefficients, $(\gamma_u, \gamma_v, \gamma_w, \gamma_\theta)$. Next, (ii) yields $\hat{\boldsymbol{u}}^\dag_v \cdot \hat{\boldsymbol{u}}_w + \hat{\theta}_v^\dag \hat{\theta}_w = \hat{\boldsymbol{u}}_v^\dag \cdot (\hat{\boldsymbol{u}} - \hat{\boldsymbol{u}}_v) + \hat{\theta}^\dag_v (\hat{\theta} - \hat{\theta}_v) = 0$, which results in $\gamma^\dag_u \hat{u} + \gamma^\dag_v \hat{v} + \gamma^\dag_w \hat{w} + \gamma^\dag_\theta \hat{\theta} = (|\gamma_u|^2 + |\gamma_v|^2 + |\gamma_w|^2 + |\gamma_\theta|^2) \hat{q}$. Comparing this expression with the definition of $\hat{q}(\hat{\boldsymbol{u}}, \hat{\theta})$ in (\ref{eq:q_lin}) yields
\begin{align}
(\gamma_u, \gamma_v, \gamma_w, \gamma_\theta) = \frac{\left( {\rm i} N \ell, - {\rm i} N k, 0, - {\rm i} \Omega m \right)}{N^2 k^2 + N^2 \ell^2 + \Omega^2 m^2} .
\end{align}
We have thus formulated a wave-vortex decomposition in the presence of a reference stream. In the same way as in the conventional case, in this definition, the wave mode does not possess potential vorticity; $\hat{q}(\hat{\boldsymbol{u}}_w, \hat{\theta}_w) = 0$.

We next derive the evolution equations of the wave mode and the vortical mode, separately. First, it is obvious that the vortical mode obeys the set of equations,
\begin{align} \label{eq:EOM_vortex}
\frac{d \hat{u}_v}{dt} = \frac{d \gamma_u}{dt} \hat{q}, \quad
\frac{d \hat{v}_v}{dt} = \frac{d \gamma_v}{dt} \hat{q}, \quad
\frac{d \hat{w}_v}{dt} = 0, \quad
\frac{d \hat{\theta}_v}{dt} = \frac{d \gamma_\theta}{dt} \hat{q} .
\end{align}
Since the vertical acceleration is identically 0 for the vortical motion, the buoyancy force should be balanced with the hydrostatic pressure represented as
\begin{align} \label{eq:hydrostatic}
\hat{p}_v = - \frac{{\rm i} N \hat{\theta}_v}{m} = - \frac{{\rm i} N \gamma_\theta \hat{q}}{m} .
\end{align}
Subtracting (\ref{eq:EOM_vortex}) from the original equation (\ref{eq:linear}) and using (\ref{eq:hydrostatic}) with $\hat{p}_w = \hat{p} - \hat{p}_v$, we also derive the equations that govern the wave mode as
\begin{subequations} \label{eq:EOM_wave}
\begin{align}
\frac{d \hat{u}_w}{dt} & = (f - \beta) \hat{v}_w - {\rm i} k \hat{p}_w + \underbrace{\left\{ \alpha \gamma_v - \frac{d \gamma_u}{dt} \right\} \hat{q}}_{F_u}  \\
\frac{d \hat{v}_w}{dt} & = (\alpha - f) \hat{u}_w - {\rm i} \ell \hat{p}_w + \underbrace{\left\{ - \beta \gamma_u - \frac{d \gamma_v}{dt} \right\} \hat{q}}_{F_v}  \\
\frac{d \hat{w}_w}{dt} & = - {\rm i} m \hat{p}_w + N \hat{\theta}_w  \\
\frac{d \hat{\theta}_w}{dt} & = - N \hat{w}_w \underbrace{- \frac{d \gamma_\theta}{dt} \hat{q}}_{F_\theta} \\
0 & = k \hat{u}_w + \ell \hat{v}_w + m \hat{w}_w .
\end{align}
\end{subequations}
We find several difference in the current system from the classical $\alpha = \beta = 0$ case. First, the vortical mode is no longer stationary because, while the potential vorticity $\hat{q}$ is unchanged, the coefficients $(\gamma_u, \gamma_v, \gamma_\theta)$ vary associated with variations in the wavevector, (\ref{eq:variation_in_k}). Second, the wave mode is not fully decoupled from the vortical mode; the terms represented as $(F_u, F_v, F_\theta)$, which are proportional to $\hat{q}$ and nonzero unless $e = 0$, work as external forces onto the wave mode. Due to these terms, even when $(\hat{\boldsymbol{u}}_w, \hat{\theta}_w)$ is initially $0$, a wave motion can be spontaneously generated from a pure vortical motion.

Let us define the wave mode energy as $\hat{E}_w = (|\hat{\boldsymbol{u}}_w|^2 + |\hat{\theta}_w|^2)/2$ and the vortical mode energy as $\hat{E}_v = (|\hat{\boldsymbol{u}}_v|^2 + |\hat{\theta}_v|^2)/2$, the sum of which coincides with the kinetic energy plus the available potential energy, $\hat{E}_w + \hat{E}_v = \hat{E}_u + \hat{E}_\theta$. The vortical mode energy is a quadratic function of the potential vorticity, $\hat{E}_v = |\hat{q}|^2/[2(N^2 k^2 + N^2 \ell^2 + \Omega^2 m^2)]$. The energy budget for each mode is described as
\begin{subequations} \label{eq:production_separate}
\begin{align}
\frac{d \hat{E}_w}{dt} & = \underbrace{(\alpha - \beta) \Re [\hat{u}_w^\dag \hat{v}_w]}_{\hat{\mathcal{P}}_w} + \underbrace{\Re[\hat{u}_w^\dag F_u + \hat{v}_w^\dag F_v + \hat{\theta}_w^\dag F_\theta]}_{\hat{\mathcal{P}}_{wv}}  \\
\frac{d \hat{E}_v}{dt} & = \underbrace{\frac{d}{dt} \left(\frac{1}{N^2 k^2 + N^2 \yo{\ell}^2 + \Omega^2 m^2}\right) \frac{|\hat{q}|^2}{2}}_{\hat{\mathcal{P}}_v} ,
\end{align}
\end{subequations}
where $\hat{\mathcal{P}}_w$ is the lateral shear production associated with the wave motion, $\hat{\mathcal{P}}_{wv}$ is the energy injection to the wave mode due to the combined effect of the reference stream and the vortical mode, and $\hat{\mathcal{P}}_v$ is the energy conversion from the reference stream to the vortical mode. Obviously, sum of these terms coincides with the total energy production rate; $\hat{\mathcal{P}}_w + \hat{\mathcal{P}}_{wv} + \hat{\mathcal{P}}_v = \hat{\mathcal{P}}$.

\subsection{Linear stability analysis}
Let us move on to the stability analysis. Although this analysis is essentially equivalent to those in previous studies, we present several new insights into this classical problem. The incompressible condition, (\ref{eq:incompressible_l}), allows us to represent the amplitude of pressure $\hat{p}$ in terms of $\hat{\boldsymbol{u}}$ and $\hat{\theta}$. Then, redefining the state variables in a vectorial form as $\boldsymbol{v} \equiv (\hat{u}, \hat{v}, \hat{w}, \hat{\theta})$, we may write the set of equations, (\ref{eq:linear}), in the form,
\begin{align} \label{eq:vA}
\frac{d \boldsymbol{v}}{dt} = \boldsymbol{A}(t) \boldsymbol{v} ,
\end{align}
where $\boldsymbol{A}(t)$ is a time-dependent 4 by 4 matrix with periodicity of $2 \pi / \omega_v \equiv T$. We then investigate the stability of the system (\ref{eq:vA}) based on Floquet theorem. In this analysis, contrary to intuition, the incompressible condition is not needed to be coupled with (\ref{eq:vA}). The reason is partly explained by \cite{mathur2014effects} and will be made more explicit below in this paper.

Stability analysis based on Floquet theorem is carried out as follows. We first prepare a set of initial conditions, $\boldsymbol{v}_1(0)= (1, 0, 0, 0), \boldsymbol{v}_2(0)= (0, 1, 0, 0), \boldsymbol{v}_3(0)= (0, 0, 1, 0), \boldsymbol{v}_4(0)= (0, 0, 0, 1)$, and then integrate (\ref{eq:vA}) for each case independently until $t=T$. The resulting state vectors are aligned to compose a 4 by 4 matrix, $\boldsymbol{M} = (\boldsymbol{v}_1, \boldsymbol{v}_2, \boldsymbol{v}_3, \boldsymbol{v}_4)$, named the monodromy matrix. We next derive the eigenvalues of $\boldsymbol{M}$ and the corresponding eigenvectors and write them as $\{ \mu_i \}$ and $\{ \tilde{\boldsymbol{w}}_i \}$, respectively. We again regard a vector $\tilde{\boldsymbol{w}}_i$ as an initial condition of (\ref{eq:vA}) and define the corresponding solution $\boldsymbol{w}_i(t)$. From the definition, this solution satisfies
\begin{align}
\boldsymbol{w}_i(T) = \boldsymbol{M} \boldsymbol{w}_i(0) = \mu_i \tilde{\boldsymbol{w}}_i .
\end{align}
We consequently understand that $\mu_i$ represents an amplification factor of the $i$th-mode state vector $\boldsymbol{w}_i$ over one period of integration of (\ref{eq:vA}). If $|\mu_i|$ is greater than unity, the $i$th mode is unstable. To measure the rate of amplification for each state vector per unit time, we shall write the growth rate as $\lambda_i = (1/T)\log|\mu_i|$.

Equation (\ref{eq:q_lin}) has an invariant, $\delta \equiv k \hat{u} + \ell \hat{v} + m \hat{w}$, which stems from the incompressibility condition imposed to eliminate the pressure variable $\hat{p}$. Since $\delta$ is a linear function of $\boldsymbol{v}$, the amplification factor of a solution $\boldsymbol{w}_i(t)$ with a nonzero $\delta$ should be unity. This mode cannot grow exponentially. That is to say, for an unstable mode, the incompressible condition, $\delta = 0$, is always satisfied. The Floquet stability analysis performed without imposing an incompressible constraint is thus rationalized. Following the same logic regarding $\hat{q}$, an unstable mode has no potential vorticity. We thus understand that pure IGWs are amplified through the AAI. In the following, for simplicity, we drop the subscript to denote the growth rate of the most unstable mode and the corresponding amplification factor and the eigen solution, namely, $\lambda \equiv \max_{i} \lambda_i$, $\mu$ and $\boldsymbol{w}(t)$, respectively.

\yo{For the stability analysis, because $k$ and $\ell$ vary composing an ellipse (\ref{eq:variation_in_k}) and the choice of the initial time is arbitrary, we can assume without loss of generality that the horizontal wavevector is initially parallel to the $x$-axis, that is, $\tilde{\ell}=0$.} The problem then involves 4 dimensionless parameters, $e$, $Ro$, $N/f$, and $\phi \equiv \arctan(\tilde{m}/\tilde{k})$, where $\phi$ represents the initial elevation angle of wavevector. The instability growth rates, $\lambda$, are computed for a range of these parameters using the 4th-order Runge-Kutta method and an eigenvalue solver in LAPACK \citep{anderson1999lapack}. For the crosscheck of our Fortran code, we have compared the results with those in \cite{mcwilliams1998fluctuation}, who adopted a different formulation, and confirmed a reasonable agreement.

\begin{figure}[t]
  \noindent\includegraphics[width=\textwidth, bb=0 0 559 213]{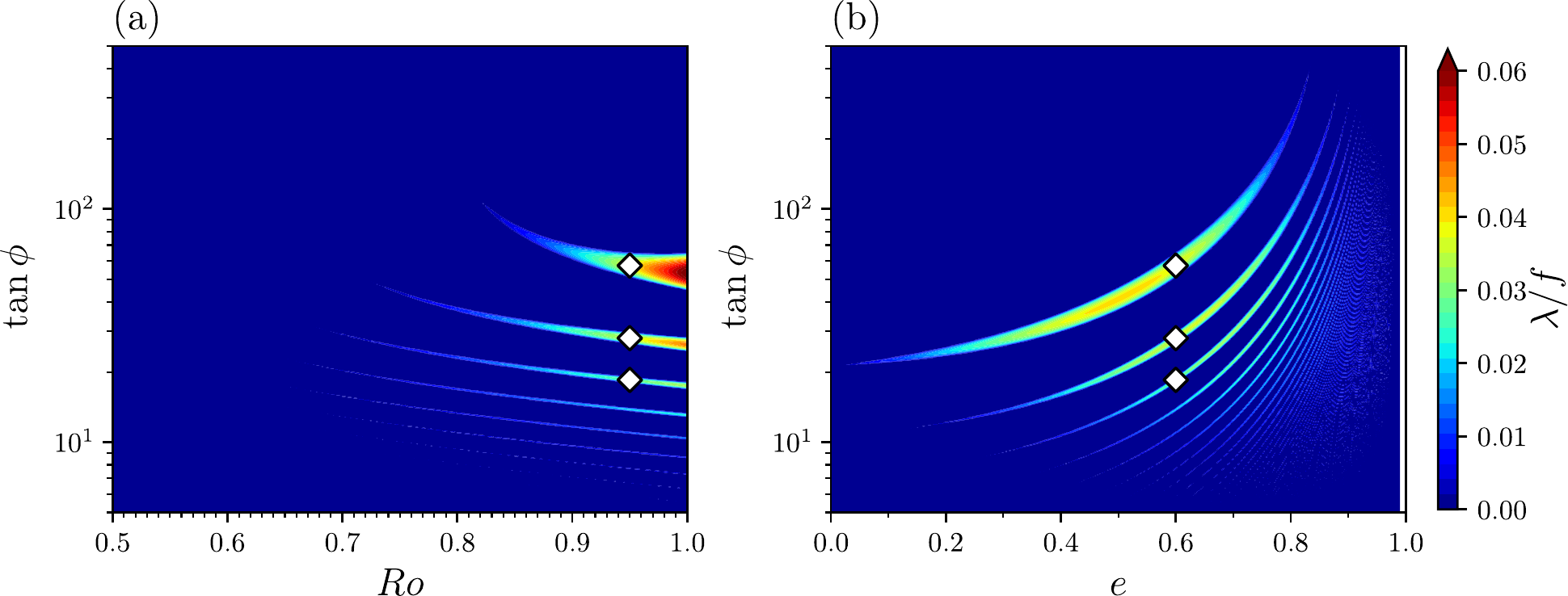}\\
  \caption{Numerically obtained instability growth rates, $\lambda$, scaled by $f$ are contoured against (a) $Ro$ and $\tan \phi$ for $e = 0.6$ and $N/f=10$ and (b) against $e$ and $\tan\phi$ for $Ro = 0.95$ and $N/f=10$. Here, $Ro$ is the Rossby number defined as the ratio of the relative vorticity to the planetary vorticity, $e$ is the ellipticity of the vortex, $\phi$ is the initial elevation angle of the wavevector, and $N/f$ is the ratio of the buoyancy frequency to the planetary vorticity. In (b), $e$ ranges from 0 to 0.99. The white diamonds indicate the parameter settings of Fig.~\ref{fig:floquet_series}.} \label{fig:growth_rate_Ro_e}
\end{figure}

\begin{figure}[t]
  \noindent\includegraphics[bb=0 0 423 807, width=0.5\textwidth]{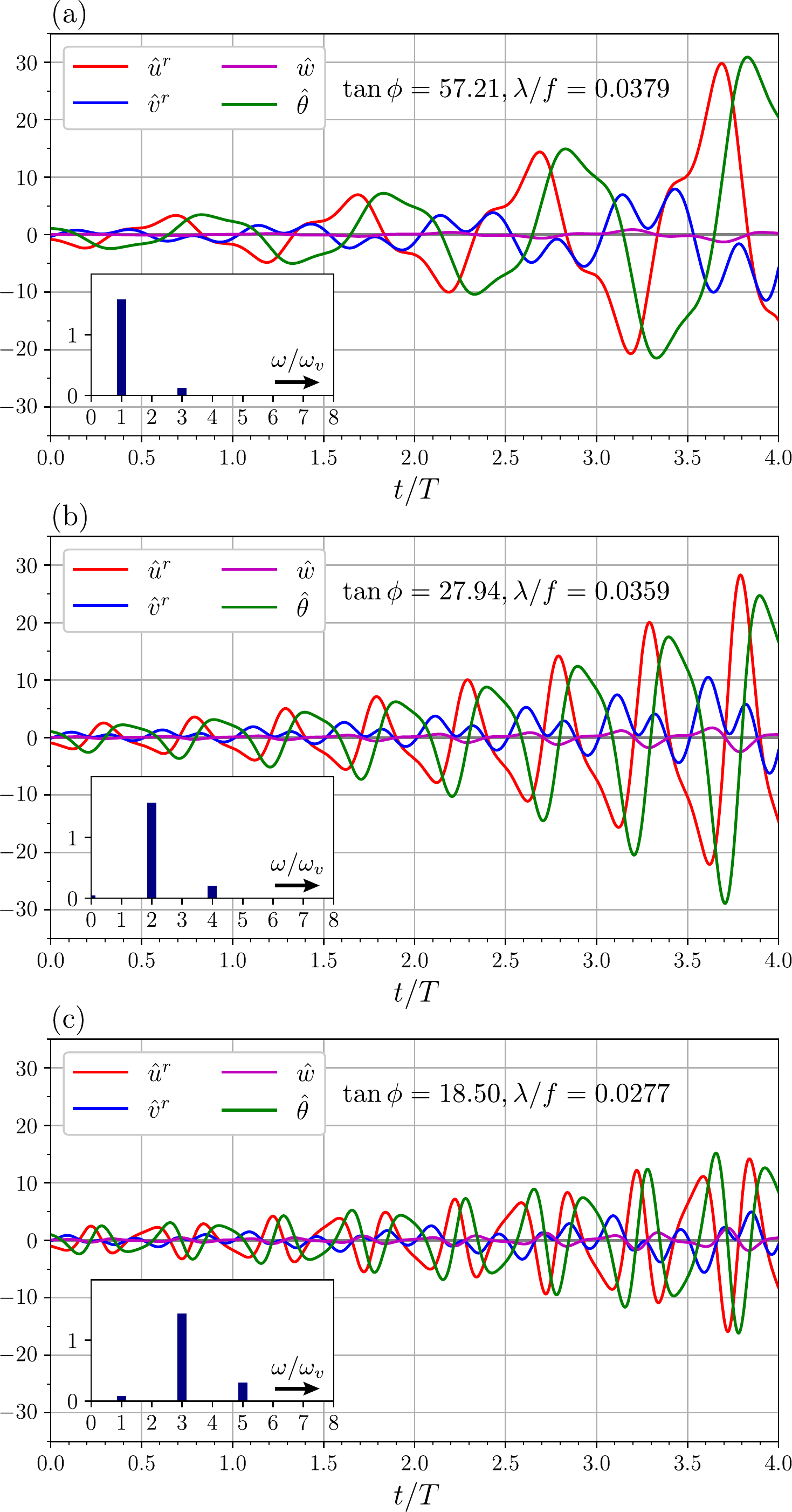}\\
  \caption{Time series of the variables, $(\hat{u}^r, \hat{v}^r, \hat{w}, \hat{\theta})$, of the unstable modes obtained from Floquet analysis for (a) $\tan \phi = 57.21$, (b) $\tan \phi = 27.94$, and (c) $\tan \phi = 18.50$, with the conditions of $Ro = 0.95, e = 0.6, N/f = 10$. The vertical axis is scaled such that the total energy, which is the sum of the kinetic and available potential energy, at the initial time is unity. The insets are the corresponding frequency spectra of the total energy density.} \label{fig:floquet_series}
\end{figure}

\begin{figure}[t]
  \noindent\includegraphics[bb=0 0 803 347, width=\textwidth]{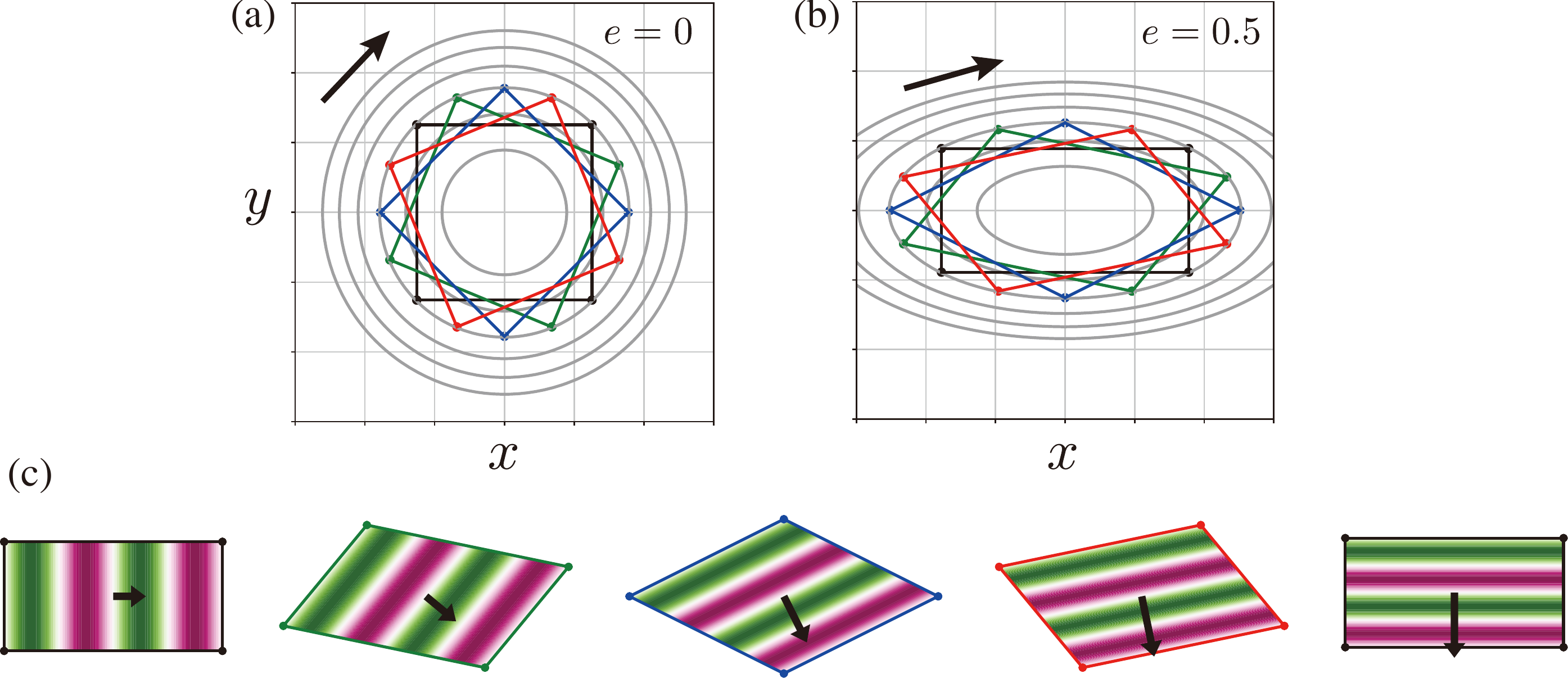}\\
  \caption{Schematic illustration of the basic mechanism of elliptic instability. We consider an initially rectangular domain embedded within an anticyclonic vortex. (a) When the vortex is circular, the domain rotates without changing its shape. (b) When the vortex is elliptic, the shape of the domain is periodically distorted; expanded in the $x$-direction while compressed in the $y$-direction. (c) Colors and arrows represent the phase and the horizontal wavevector of a plane wave disturbance in the initially rectangular domain within the elliptic vortex in (b). The length of the wavevector varies over time; lengthened and shortened twice during one cycle of rotation. If the natural frequency of this wave matches with the vortex rotation rate, parametric resonance occurs resulting in exponential wave amplification.} \label{fig:parametric_resonance}
\end{figure}

Figure~\ref{fig:growth_rate_Ro_e}a demonstrates the computed instability growth rates against $Ro$ and $\tan \phi$ for the case of $(N/f,e)= (10, 0.6)$. Several wedge-shaped unstable regions extend from larger to smaller $Ro$ directions. To qualitatively distinguish the wedges, we analyze the temporal character of the growing mode in further detail. The method basically follows that of \cite{onuki2019instabilities}; the eigen solution is reduced by a time-dependent amplification factor to compose a $T$-periodic function, $\boldsymbol{w}_p = \boldsymbol{w} \mu^{- t/T}$, and $\boldsymbol{w}_p(t)$ is expanded into a Fourier series, $\boldsymbol{w}_p = \sum_j \boldsymbol{W}_j {\rm e}^{{\rm i} j \omega_v t}$, to derive a frequency spectrum. Several cautions are required in this analysis. First, the actual frequency of the eigen solution, $\boldsymbol{w}$, differs from that of $\boldsymbol{w}_p(t)$ because of the argument of $\mu$. The frequency axis should be accordingly represented as $\omega = j \omega_v + (1/T) \Im [\log \mu] (j = 0, \pm 1, \pm 2, \ldots)$, where $\Im$ represents taking the imaginary part. Second, the frequency spectrum derived using the Fourier coefficients, $(\hat{\boldsymbol{u}}, \hat{\theta})$, deviates from that in physical space using $(\boldsymbol{u}, \theta)$ because the variation in wavenumber $\boldsymbol{k}(t)$ is not taken into account. This difference corresponds to the Doppler effect caused by the elliptic stream. We thus interpret $\omega$ used here as an intrinsic frequency measured at a frame moving with the background stream. Third, in the moving frame rotating with the background stream, the reference direction of disturbance velocity also rotates at a mean rate of $\omega_v$ around the vertical axis. To incorporate this effect into the spectrum, we redefine the horizontal velocity vector as
\begin{align} \label{eq:velocity_rotate}
(\hat{u}^r, \hat{v}^r) \equiv (\hat{u} \cos \omega_v t - \hat{v} \sin \omega_v t, \hat{u} \sin \omega_v t + \hat{v} \cos \omega_v t)
\end{align}
and Fourier expand $(\hat{u}^r, \hat{v}^r)$ instead of $(\hat{u}, \hat{v})$. Figure~\ref{fig:floquet_series} shows the eigen solutions of the unstable modes and the corresponding energy spectra obtained for $(N/f, Ro, e) = (10, 0.95, 0.6)$, and $\tan\phi =$ 57.21, 27.94, and 18.50, within three different wedges. For the most unstable mode ($\tan\phi = 57.21$), the energy peak is located at $\omega = \omega_v$. This result is explained from the basic mechanism of parametric excitation (Fig.~\ref{fig:parametric_resonance}); an IGW with its natural frequency $\omega_v$ is resonantly excited via the periodic stretching of the horizontal wavenumber, $k_H \equiv \sqrt{k^2 + \ell^2}$. For other branches, the peak frequency is always located at a multiple of the vortex frequency and shifts higher as $\phi$ decreases, as inferred from the dispersion relationships of IGWs. We can thus distinguish the unstable regions in parameter space in terms of the intrinsic wave frequency divided by the vortex frequency.
Wedge-shaped unstable regions are also visible in Fig.~\ref{fig:growth_rate_Ro_e}b at $e < 0.8$. The gravest wedge that extends to close to $e=0$ again corresponds to the region with peak frequency at $\omega_v$. For a large $e$ area, width of the wedges gets much thinner, and the unstable regions are no longer distinguishable. Consequently, IGWs of an almost continuous frequency spectrum will be excited.

\begin{figure}[t]
  \noindent\includegraphics[bb=0 0 567 217, width=\textwidth]{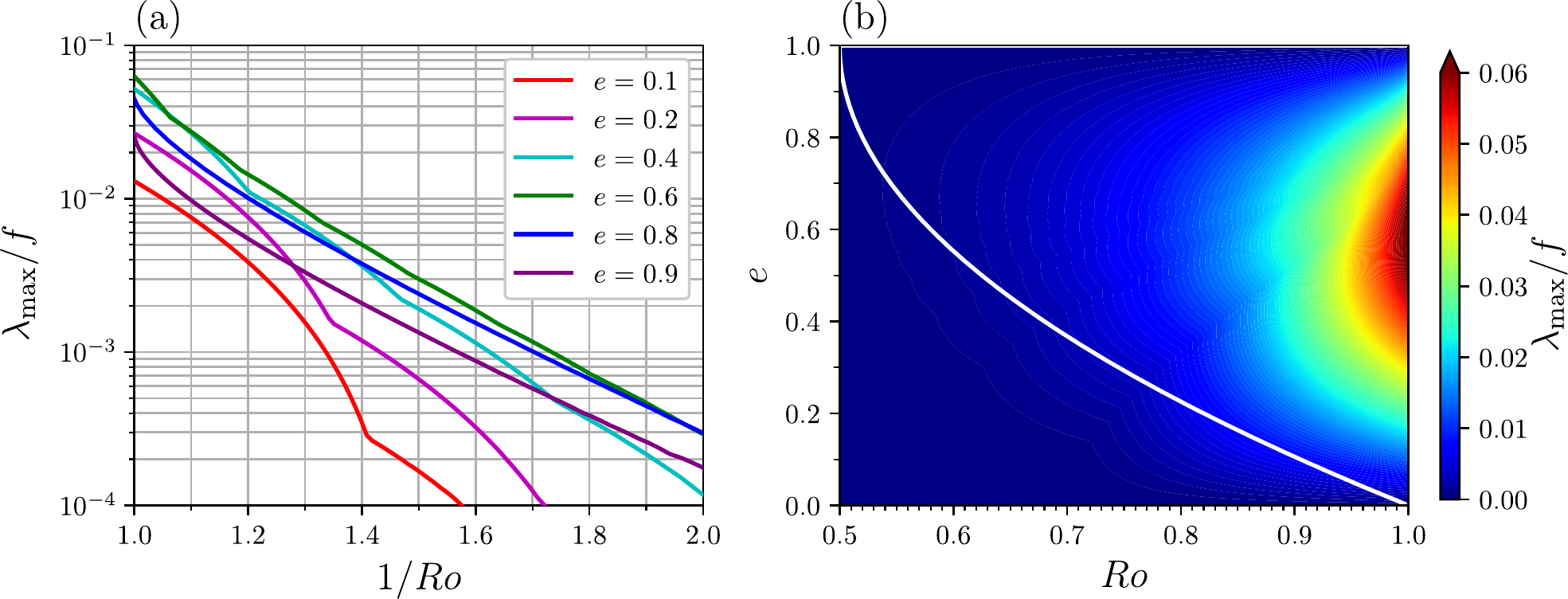}\\
  \caption{(a) The maximum instability growth rate in wavenumber space, $\lambda_{\rm max} \equiv {\rm max}_{\phi}\lambda$, scaled by $f$ plotted against the inverse Rossby number, $1/Ro$, for several values of ellipticity, $e$. (b) Contour plot of $\lambda_{\rm max}/f$ against $Ro$ and $e$. The white curve denotes the line of $A - S = 0$, where $A$ is the absolute vorticity and $S$ is the strain rate. Their precise definitions are given in (\ref{eq:A_S}). In both plots, we adopt $N/f=10$.} \label{fig:growth_rate}
\end{figure}

Figure~\ref{fig:growth_rate}a shows the maximum growth rate in wavenumber space, or $\lambda_{\rm max} = \max_\phi \lambda$, as a function of $1 / Ro$. Note that the vertical axis is a logarithmic scale while the horizontal axis is a linear scale. Roughly, $\lambda_{\rm max} \sim {\rm e}^{-C/Ro}$ predicted from the asymptotic analysis \citep{aspden2009elliptical} is confirmed. In these plots, we notice that the curve slope suddenly changes at several points. This correspond\yo{s} to the exchange in the maximum growth rates from one wedge-shaped unstable region to another in the $Ro$-$\tan\phi$ plane (Fig.~\ref{fig:growth_rate_Ro_e}a). Figure~\ref{fig:growth_rate}b also shows a dependence of $\lambda_{\rm max}$ against $e$ as well as $Ro$. It is natural that $\lambda_{\rm max} = 0$ at $e=0$ because the energy production term $\mathcal{P}$ defined as (\ref{eq:energy_production}) vanishes at a circular vortex case, $\alpha = \beta$. Besides, we find that $\lambda_{\rm max}$ is maximum at around $e = 0.6$ and significantly decreases when $e$ approaches 1. This result indicates that an extremely elongated vortex does not induce exponential growth of inertia-gravity wave energy. \cite{mcwilliams1998fluctuation} pointed out that, in a homogeneous parallel shear flow, the disturbance wave amplitude remains finite even for an asymptotically large time. Therefore, it is reasonable that the most effective instability occurs somewhere within $0 < e < 1$. Classically, the difference between the absolute vorticity $A$ and the strain rate $S$ has been used for the criterion of the occurrence of AAI. In the present case,
\begin{align} \label{eq:A_S}
\frac{A - S}{f} = 1 - Ro - \frac{1 - (1 - e)^2}{1 + (1 - e)^2} Ro < 0
\end{align}
would be regarded as the possibly unstable region. In Fig.~\ref{fig:growth_rate}b, it is true that $A - S = 0$ well follows the contour of $\lambda_{\rm max} \sim 0.01 f$ at $e \leq 0.25$. However, for a larger $e$, since the system is again stabilized, this criterion is virtually meaningless.

In the present linear analysis, we have employed an inviscid system. As a result, the computed instability growth rate does not depend on the wavenumber length. Only the initial elevation angle of wavevector matters. Now, we take the terms of viscosity and diffusivity into account. For a particular case when viscosity and diffusivity coincide, $\nu_u = \nu_\rho = \nu$, the viscous solution, $\boldsymbol{w}^\nu_i$, is immediately derived from the inviscid solution, $\boldsymbol{w}_i$, as $\boldsymbol{w}_i^\nu (t) = \boldsymbol{w}_i(t) \exp \left( - \int_0^t \nu |\boldsymbol{k}(t')|^2 dt' \right)$. As a result, the viscous growth rate, $\lambda^v$, is explicitly derived as
\begin{align}
\lambda^\nu = \lambda - \nu \tilde{k}^2 \left( \frac{1 + r^2}{2} + \tan^2 \phi \right) .
\end{align}
This expression tells us that the growth rate is a monotonic decreasing function of the initial wavenumber length. Although the current \yo{homogeneous base flow condition} does not allow the existence of the lower bound in wavenumber length, in practice, it can be said that the possible largest-scale mode primarily grows. The actual shape and size of this primary mode are determined by the geometrical conditions, such as a diameter or a thickness of a reference vortex. The local stability analysis is incapable of coping with this scale-selection problem. However, this simple analysis offers fruitful insights into the parameter dependence and temporal characteristics of the growing inertia-gravity wave disturbance within a sheared reference stream. The information obtained here are utilized for the model \yo{setup} and data interpretation in the fully nonlinear simulations.

\begin{figure}[t]
  \noindent\includegraphics[bb=0 0 426 494, width=0.5\textwidth]{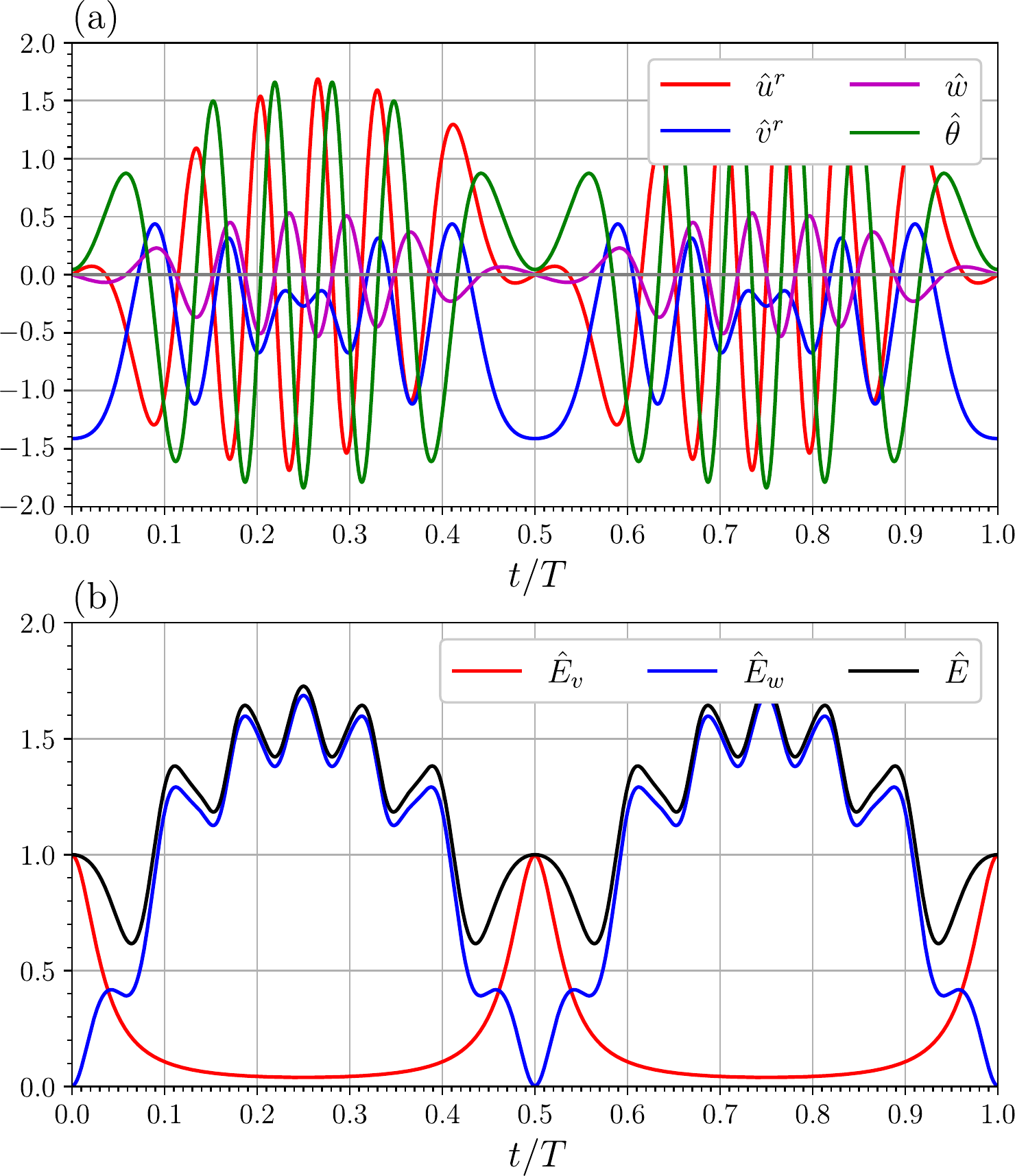}\\
  \caption{(a) Time series of the variables, $(\hat{u}^r, \hat{v}^r, \hat{w}, \hat{\theta})$, of the neutrally stable mode for $Ro = 0.98, e = 0.8, N/f = 10$, and $\tan \phi = 15.59$. (b) Energy of the wave mode ($\hat{E}_v$, red), the vortical mode ($\hat{E}_w$, blue), and their sum ($\hat{E}$, black).} \label{fig:floquet_neutral}
\end{figure}

Before concluding this section, we further mention neutrally stable modes. In addition to the unstable IGW solution, we can find an eigenvector of $\bm{M}$ that has $\hat{q} \neq 0$ and $\delta = 0$. Although this solution is characterized by the potential vorticity, it is not a pure vortical mode. As clarified in (\ref{eq:EOM_wave}), any finite $\hat{q}$ drives a wave mode. In particular when $e$ is large and $Ro \sim 1$, the wave mode amplitude forced by the vortical mode can become huge \citep{mcwilliams1998fluctuation} despite the amplification factor derived from Floquet analysis being unity. For example, Fig.~\ref{fig:floquet_neutral} shows the time series of the variables as well as the wave and vortical mode energy for the neutrally stable mode of the $(N/f, Ro, e, \tan\phi) = (10, 0.98, 0.8, 15.59)$ case. Although most of energy is contained in the vortical mode at the initial time, the wave mode energy is transiently enhanced during one cycle of \yo{vortex} rotation. This study regards this wave generation mechanism as another possible energy cascade process.

\section{Nonlinear simulations}
\subsection{Model configuration}
\begin{figure}[t]
  \noindent\includegraphics[bb=0 0 388 465, width=0.5\textwidth]{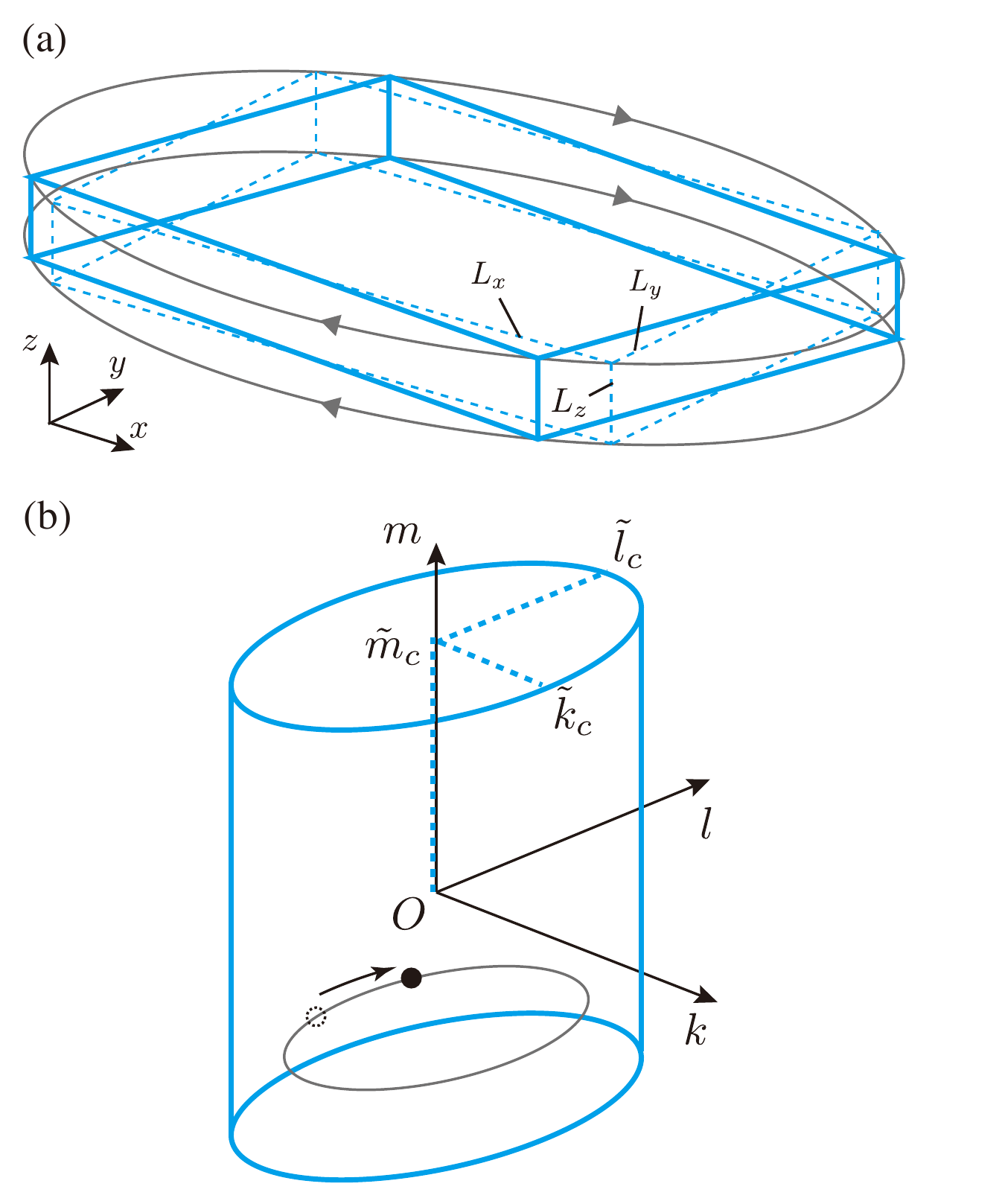}\\
  \caption{(a) Schematic view of the numerical model in physical space. The calculation domain is initially a rectangular represented as dashed blue lines. The model geometry is specified by the edge lengths of this rectangular, $L_x$, $L_y$ and $L_z$. The domain rotates following the reference streamlines indicated by gray ellipses while being distorted as represented by blue solid lines. (b) Schematic view in wavenumber space. The calculation domain is an elliptic cylinder whose geometry is specified by the truncation wavenumbers, $\tilde{k}_c$, \yo{$\tilde{\ell}_c$} and $\tilde{m}_c$. Although the overall structure of the domain is time-invariant, each element that composes the domain moves following an ellipse similar to the edges of the cylinder.} \label{fig:model_concept}
\end{figure}

To investigate the saturation and dissipation processes of growing \yo{IGWs} via AAI, we carry out fully nonlinear numerical simulations. The Fourier integral is now replaced by a Fourier series expansion. To make the formulation simpler, we introduce a moving coordinate in physical space, $\boldsymbol{X}(\boldsymbol{x}, t)=(X, Y, Z)$. This coordinate initially coincides with $\boldsymbol{x}$ but moves following the reference stream; i.e., $\boldsymbol{X}(0) = \boldsymbol{x}$ and $D_t \boldsymbol{X} = 0$. Accordingly, we derive
\begin{align}
X = x \cos \omega_v t - ry \sin \omega_v t, \quad Y = \frac{x}{r} \sin \omega_v t + y \cos \omega_v t, \quad Z = z .
\end{align}
The calculation domain is a rectangular fixed in the $\boldsymbol{X}$ coordinate, namely, $0 \leq X \leq L_x$, $0 \leq Y \leq L_y$, and $0 \leq Z \leq L_z$, which corresponds to a rotating parallelepiped in the $\boldsymbol{x}$ coordinate (Fig.~\ref{fig:model_concept}a). Imposing a triple periodic boundary condition with respect to $\boldsymbol{X}$, a Fourier series expansion is defined as $\boldsymbol{u} = \sum_{\tilde{\boldsymbol{k}}} \hat{\boldsymbol{u}}_{\tilde{\boldsymbol{k}}} {\rm e}^{{\rm i} \tilde{\boldsymbol{k}} \cdot \boldsymbol{X}}$, where the wavevector, $\tilde{\boldsymbol{k}} = (\tilde{k}, \tilde{\ell}, \tilde{m})$, is discretized as $\tilde{k} \in (-N_x/2, \cdots , N_x/2-1)(2 \pi / L_x)$, $\tilde{\ell} \in (-N_y/2, \cdots , N_y/2-1) (2 \pi / L_y)$, $\tilde{m} \in (-N_z/2, \cdots , N_z/2-1) (2 \pi / L_z)$. Notably, this formulation is equivalent to the discretized form of the time-dependent Fourier integral (\ref{eq:Fourier_time_dependent}), and efficiently computed using a fast Fourier transform algorithm. Numerical integration of the governing equations (\ref{eq:governing_equations_k}) basically follows \cite{chung2012direct}. The pressure terms are evaluated imposing the incompressible condition. The viscous and diffusive terms are analytically treated. The quadratic terms are calculated in physical space and dealiased using the 2/3 rule. For the temporal evolution, we employ the 3rd-order Runge-Kutta scheme of \cite{spalart1991spectral}.

To specify the domain size, We first suppose that the domain shape restores to the initial shape every $90^\circ$ rotation by imposing $L_x/L_y = r$. Next, we fix the vertical length as $L_z = 200$ m and arrange the vertical aspect ratio as $L_x/L_z = \tan \phi_{\rm max}$, where $\phi_{\rm max}$ is the elevation angle of the wavevector at which the instability growth rate is maximum, so that a wave mode of the largest wavelength components, $\tilde{\boldsymbol{k}} = (\pm 2\pi / L_x, 0, \pm 2\pi / L_z)$ or $(0, \pm 2\pi / L_y, \pm 2\pi / L_z)$, grows fastest. In the following, we shall call these components as the primary mode.

The truncation wavenumbers are now $\tilde{k}_c = [N_x / 3](2\pi/L_x), \tilde{\ell}_c = [N_y / 3](2\pi/L_y)$, and $\tilde{m}_c = [N_z / 3](2\pi/L_z)$, where $[]$ is the floor function. We eliminate the horizontal wavevectors outside an ellipse $(\tilde{k} / \tilde{k}_c)^2 + (\tilde{\ell} / \tilde{\ell}_c)^2 = 1$, so that the effective wavenumbers compose an elliptic cylinder. We further assume that $N_x = N_y$, which makes the effective genuine wavenumbers, $\boldsymbol{k}(\tilde{\boldsymbol{k}}, t)$, also compose the congruent elliptic cylinder that is time-invariant. Consequently, the calculation domain is distorted in physical space but fixed in wavenumber space (Fig.~\ref{fig:model_concept}b).

In our simulations, the number of \yo{grid points} is insufficient to resolve the Kolmogorov scale. To make the viscous and diffusive terms as small as possible at low wavenumbers while avoiding energy accumulation near the truncation wavenumber, we incorporate a subgrid model that modifies the viscosity and diffusivity as $\nu_u = \nu_{u0} + \nu_{ue}(\boldsymbol{k})$ and $\nu_\rho = \nu_{\rho 0} + \nu_{\rho e}(\boldsymbol{k})$. Functional forms of the wavenumber-dependent parts representing the eddy viscosity and diffusivity, $\nu_{ue}$ and $\nu_{\rho e}$, are specified in Appendix A.

\renewcommand{\arraystretch}{1.2}
\begin{table}[t]
\caption{Experimental parameters for the nonlinear numerical simulations. Here, \yo{$Ro$ and $e$ are Rossby number and ellipticity defined for the reference flow, $N/f$ is the buoyancy frequency divided by the planetary vorticity, and} $\theta_{max}$ and $\lambda^\nu_{max}$ are the initial elevation angle of the wavevector and the instability growth rate of the primary mode, respectively. \yo{In all the experiments, the Coriolis parameter is fixed as $f = 1.0 \times 10^{-4}$ rad/s.} \yo{As to the geometrical conditions}, the vertical length of the domain is fixed as $L_z = 200$ m, while the horizontal lengths at the initial time are varied depending on the parameter setting as $L_x = \tan \theta_{max} L_z$ and $L_y = L_x / r$, where $r = 1 / (1 - e)$. The number of \yo{grid points, $N_x \times N_y \times N_z$,} is $3072 \times 3072 \times 384$ for the run \textbf{Id 1}, and $768 \times 768 \times 192$ for the other runs. The viscosity and diffusivity are fixed for \textbf{Id 1} and varied for the others using the subgrid model (Appendix A). The fixed parts of the viscosity and diffusivity, $\nu_{u 0}$ and $\nu_{\rho 0}$, are $1.0 \times 10^{-4}$ [m$^2$/s] for \textbf{Id 1, 2}, and $1.0 \times 10^{-6}$ [m$^2$/s] for the others. To make explicit the exceptional viscosity conditions, asterisks are placed on ID 1 and 2 In the table.}\label{t1}
\begin{center}
\begin{tabular}{cccccccccccc}
\hline\hline
\textbf{Id} & $Ro$ & $e$ & $N/f$ & $\tan\theta_{max}$ & $\lambda^\nu_{max}/f$ & \textbf{Id} & $Ro$ & $e$ & $N/f$ & $\tan\theta_{max}$ & $\lambda^\nu_{max}/f$ \\
\hline
$\boldsymbol{1}^\ast$ & 1 & 0.6 & 3 & 15.80 & 0.0613 & $\boldsymbol{2}^{\ast\ast}$ & 1 & 0.6 & 3 & 15.80 & 0.0613 \\
\textbf{3} & 1 & 0.6 & 3 & 15.80 & 0.0623 & \textbf{4} & 1 & 0.6 & 6 & 31.78 & 0.0628 \\
\textbf{5} & 1 & 0.6 & 8 & 42.42 & 0.0629 & \textbf{6} & 1 & 0.6 & 10 & 53.04 & 0.0629 \\
\textbf{7} & 1 & 0.8 & 6 & 105.6 & 0.0449 & \textbf{8} & 1 & 0.7 & 8 & 67.87 & 0.0580 \\
\textbf{9} & 1 & 0.5 & 10 & 38.43 & 0.0603 & \textbf{10} & 1 & 0.4 & 5 & 15.25 & 0.0517 \\
\textbf{11} & 1 & 0.3 & 7 & 18.16 & 0.0401 & \textbf{12} & 1 & 0.2 & 4 & 9.178 & 0.0264 \\
\textbf{13} & 1 & 0.5 & 2 & 7.526 & 0.0582 & \textbf{14} & 1 & 0.7 & 4 & 33.86 & 0.0579 \\
\textbf{15} & 1 & 0.4 & 3 & 9.084 & 0.0510 & \textbf{16} & 1 & 0.55 & 3 & 13.25 & 0.0617 \\
\textbf{17} & 1 & 0.25 & 2 & 4.737 & 0.0317 & \textbf{18} & 1 & 0.3 & 2 & 5.054 & 0.0381 \\
\textbf{19} & 1 & 0.5 & 4 & 15.30 & 0.0598 & \textbf{20} & 1 & 0.7 & 3 & 25.34 & 0.0577 \\
\textbf{21} & 1 & 0.75 & 2 & 23.20 & 0.0520 & \textbf{22} & 1 & 0.5 & 3 & 11.43 & 0.0594 \\
\textbf{23} & 0.99 & 0.6 & 2 & 10.56 & 0.0561 & \textbf{24} & 0.99 & 0.6 & 4 & 21.39 & 0.0571 \\
\textbf{25} & 0.99 & 0.7 & 3 & 25.67 & 0.0504 & \textbf{26} & 0.99 & 0.5 & 3 & 11.56 & 0.0553 \\
\textbf{27} & 0.99 & 0.5 & 2 & 7.610 & 0.0542 & \textbf{28} & 0.98 & 0.4 & 3 & 9.290 & 0.0451 \\
\textbf{29} & 0.98 & 0.35 & 2 & 5.581 & 0.0394 & \textbf{30} & 0.98 & 0.7 & 4 & 34.89 & 0.0438 \\
\textbf{31} & 0.98 & 0.8 & 2 & 11.15 & 0.0329 & \textbf{32} & 0.98 & 0.65 & 3 & 20.08 & 0.0489 \\
\textbf{33} & 0.97 & 0.4 & 5 & 15.79 & 0.0428 & \textbf{34} & 0.97 & 0.6 & 3 & 16.45 & 0.0468 \\
\textbf{35} & 0.97 & 0.55 & 3 & 13.76 & 0.0481 & \textbf{36} & 0.97 & 0.5 & 4 & 15.87 & 0.0479 \\
\textbf{37} & 0.97 & 0.7 & 3 & 12.82 & 0.0399 & \textbf{38} & 0.97 & 0.55 & 2 & 9.075 & 0.0474 \\
\textbf{39} & 0.96 & 0.5 & 2 & 7.922 & 0.0431 & \textbf{40} & 0.96 & 0.6 & 3 & 16.73 & 0.0421 \\
\textbf{41} & 0.96 & 0.7 & 2 & 8.448 & 0.0360 & \textbf{42} & 0.96 & 0.65 & 2 & 13.78 & 0.0378 \\
\textbf{43} & 0.95 & 0.6 & 6 & 34.29 & 0.0379 & \textbf{44} & 0.95 & 0.1 & 6 & 13.45 & 0.00983 \\
\textbf{45} & 0.95 & 0.4 & 2 & 6.359 & 0.0359 & \textbf{46} & 0.94 & 0.2 & 5 & 12.36 & 0.0188 \\
\textbf{47} & 0.93 & 0.3 & 8 & 22.68 & 0.0259 & \textbf{48} & 0.92 & 0.1 & 6 & 14.03 & 0.00811 \\
\textbf{49} & 0.91 & 0.2 & 3 & 7.71 & 0.0150 & \textbf{50} & 0.9 & 0.6 & 6 & 17.99 & 0.0255 \\
\textbf{51} & 0.9 & 0.1 & 3 & 7.184 & 0.00691 & \textbf{52} & 0.9 & 0.1 & 10 & 24.20 & 0.00704 \\
\textbf{53} & 0.9 & 0.2 & 5 & 13.16 & 0.0141 & \textbf{54} & 0.89 & 0.1 & 6 & 14.75 & 0.00650 \\
\textbf{55} & 0.88 & 0.3 & 8 & 24.81 & 0.0172 & \textbf{56} & 0.87 & 0.2 & 4 & 11.13 & 0.0109 \\
\textbf{57} & 0.85 & 0.6 & 6 & 19.73 & 0.0165 & \textbf{58} & 0.8 & 0.6 & 6 & 13.76 & 0.0111 \\
\hline
\end{tabular}
\end{center}
\end{table}

Detailed parameter settings are summarized in Table 1. For most of simulations, the background buoyancy frequency, $N$, is smaller than the oceanic typical values. We note in Appendix B that the instability growth rate hardly depends on the stratification as far as $N > f$ is satisfied, but the wavevector elevation angle of the most unstable mode increases with~$N$. Therefore, if we made $N$ greater, the computation domain would be elongated horizontally. As a consequence, the number of horizontal grid points required to resolve wave breaking would become huge that is beyond our computer resources.

The experimental procedure is the same for all the simulations. We give a small noise only to the wave mode at the smallest wavenumbers, $-2\pi / L_x \leq \tilde{k} \leq 2\pi / L_x, -2\pi / L_y \leq \tilde{\ell} \leq 2\pi / L_y$, and $-2\pi / L_z \leq \tilde{m} \leq 2\pi / L_z$ at the initial time, and let the system evolve without any external forcing term.

\subsection{Case I: strong instability}
First, we focus on the experiment of $(N/f, Ro, e) = (3, 1.0, 0.6)$, when the instability growth rate is \yo{quite large}. We carried out three simulations for this parameter setting but different resolutions or viscosity conditions (\textbf{Id 1, 2, 3} in Table 1). Among these, \textbf{Id 1, 2} are conducted with relatively high $\nu_{u 0}$ and $\nu_{\rho 0}$ conditions for a calibration purpose (Appendix A) while \textbf{Id 3} is conducted with low $\nu_{u 0}$ and $\nu_{\rho 0}$ for the use of quantitative discussion in Section 4. Here, we demonstrate the results of \textbf{Id 2} as the representative.

\begin{figure}[t]
  \noindent\includegraphics[bb=0 0 406 485, width=0.5\textwidth]{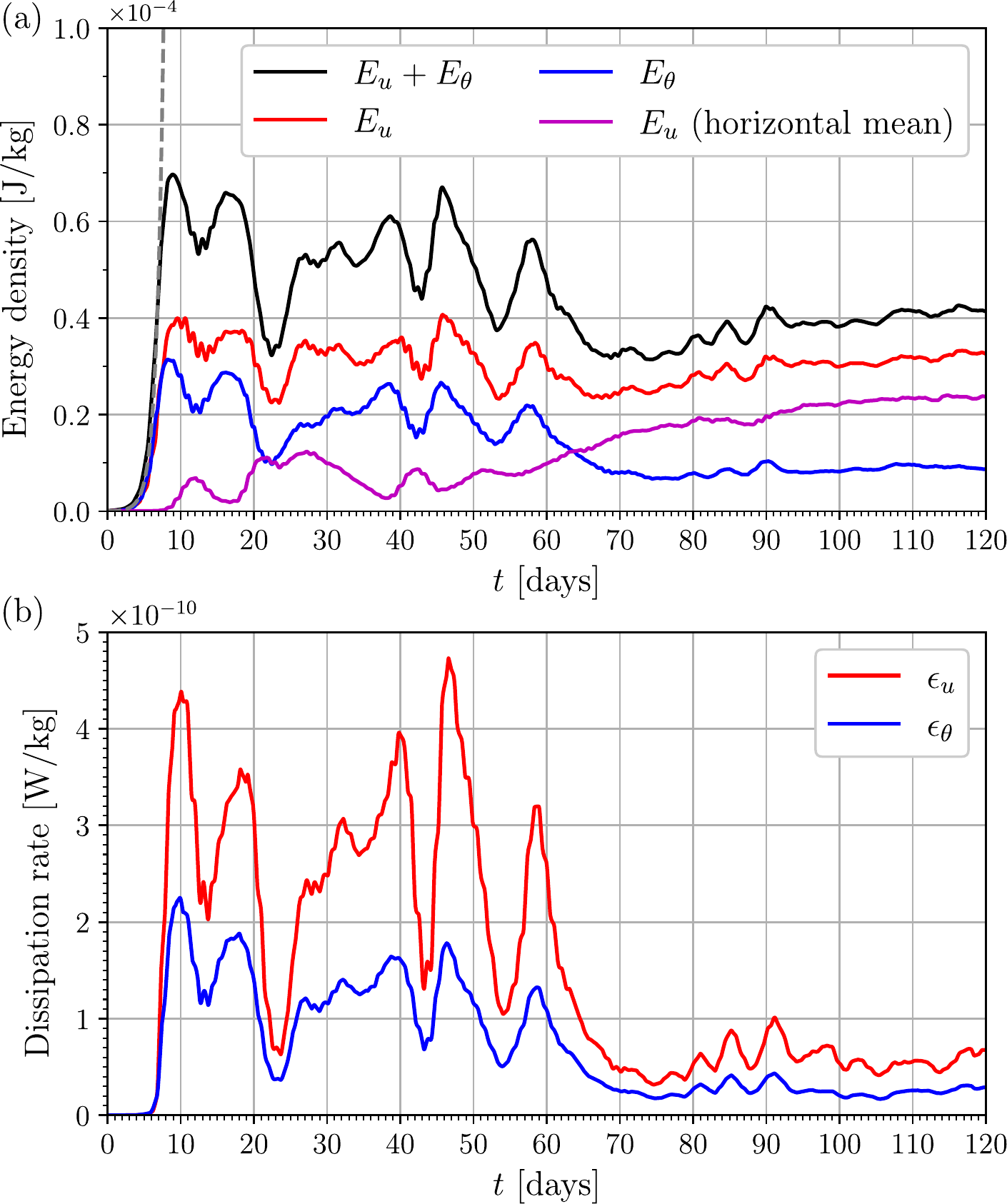}\\
  \caption{(a) Time series of the kinetic energy, available potential energy, their sum, and the kinetic energy of the horizontal mean ($k = \yo{\ell} = 0$) components for the run \textbf{Id 2}. Data are moving averaged over one period of vortex rotation. The grey broken line indicates the theoretically estimated energy growth rate. (b) Those of the kinetic energy and available potential energy dissipation rates.} \label{fig:time_series_case_1}
\end{figure}

\begin{figure}[t]
  \noindent\includegraphics[bb=0 0 508 508, width=\textwidth]{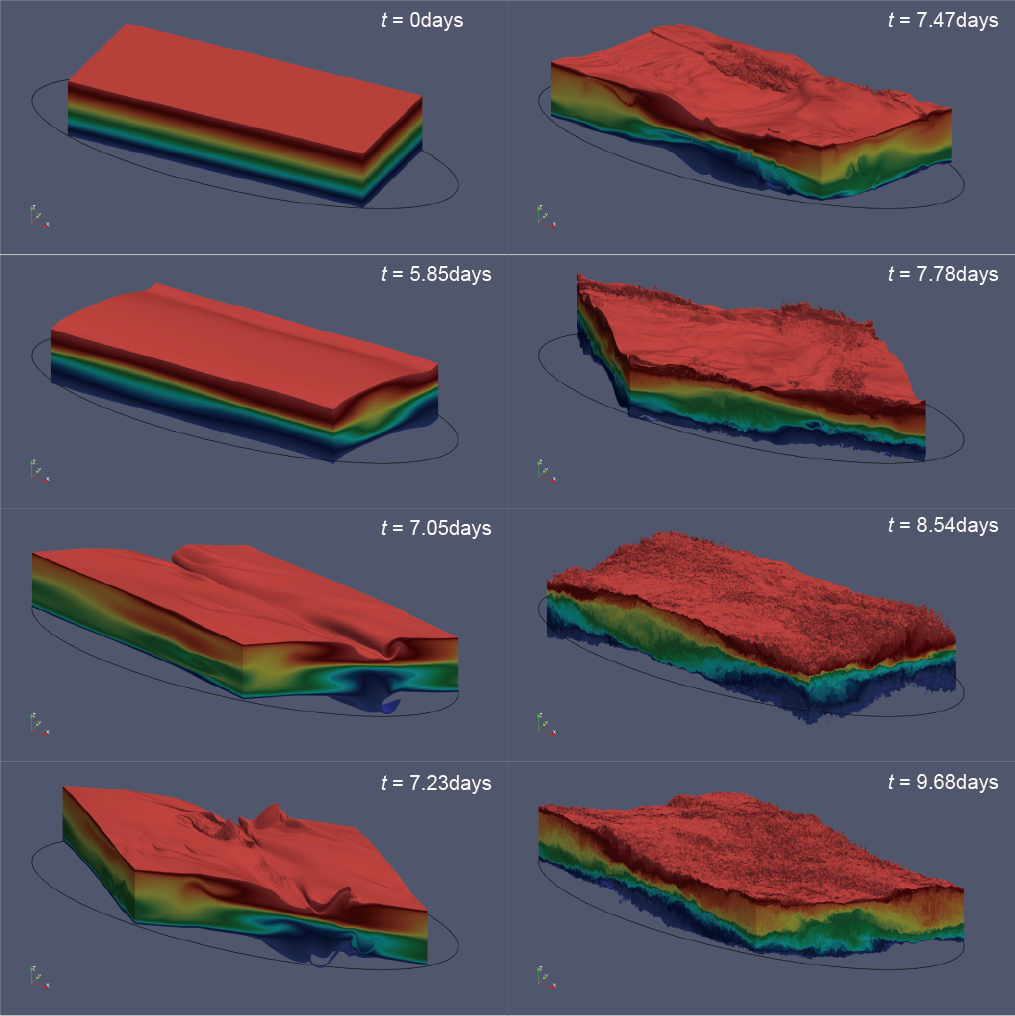}\\
  \caption{Evolution of the density field for the run \textbf{Id 2}. Displacements of the isopycnal surfaces and the stratification structures at the lateral faces are visualized. The black ellipse indicates the streamline of the background vortex flow. The vertical scale is magnified three times.} \label{fig:Rho_case_1}
\end{figure}

\begin{figure}[t]
  \noindent\includegraphics[bb=0 0 1556 1244, width=0.5\textwidth]{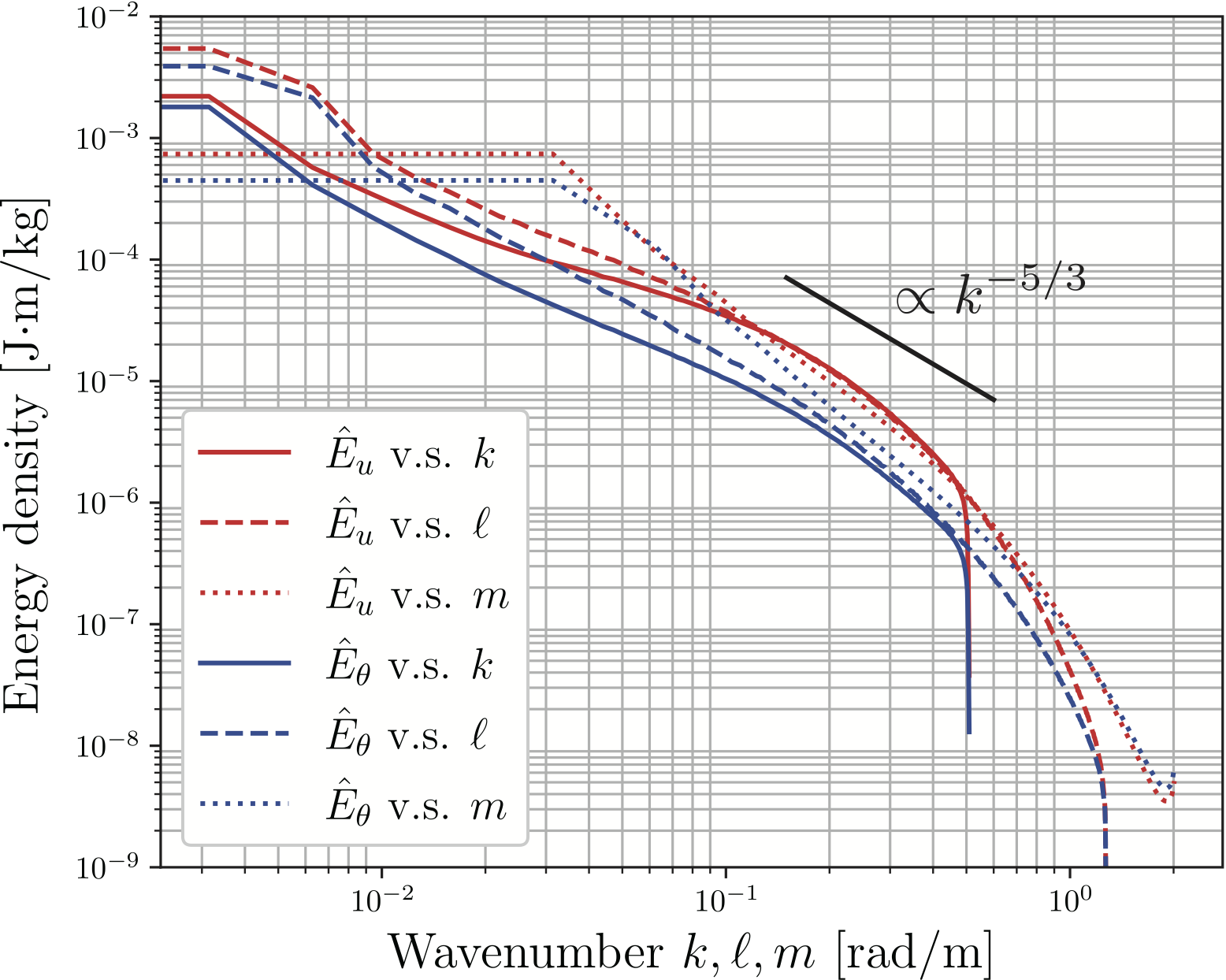}\\
  \caption{Kinetic and available potential energy spectra for the run \textbf{Id 2}. Data are averaged over 11 days $\leq t <$ 13 days and projected onto the $k$-, \yo{$\ell$}-, or $m$-axis to create 1d plots.} \label{fig:energy_spectrum}
\end{figure}

Figure~\ref{fig:time_series_case_1}a is a time series of the kinetic and available potential energy. In the initial stage of exponential energy growth, the growth rate evidently coincides with that predicted from linear stability analysis. Around $t \sim 8$ days, the energy ceases growing and, at the same time, the dissipation rates of the kinetic and available potential energy are abruptly enhanced. The spatial structure of the amplified internal wave and its breaking process are illustrated in Fig.~\ref{fig:Rho_case_1} by coloring the density field before and after the timing of energy saturation. At $t \sim 7$ days, a wave crest is significantly stretched and, consequently, dense water is placed on liter water resulting in the occurrence of gravitational instability that produces small-scale turbulent fluctuations. Figure~\ref{fig:energy_spectrum} shows the kinetic and available potential energy spectra for each direction in a turbulent state. In the low-wavenumber region, $\lesssim 0.1$ rad/m, the energy level in the vertical spectrum overwhelms that of the horizontal spectra, but in the higher wavenumber region, energy levels almost coincide for the three directions. According to the standard scaling estimates for a stratified fluid \citep{carnevale2001buoyancy,kimura2012energy}, it is known that an energy spectrum follows $\hat{E}_{\yo{u}}(\kappa) \sim c N^2 \kappa^{-3}$ for low wavenumbers and $\hat{E}_{\yo{u}}(\kappa) \sim C_K \epsilon_u^{2/3} \kappa^{-5/3}$ for high wavenumbers. Here, $\kappa$ is a wavenumber, $c$ is a constant that differs for the horizontal and vertical directions, and $C_K \sim 1.5$ is the Kolmogorov constant. Consequently, transition in the spectral slope occurs at $\kappa = (c / C_K)^{3/4} (N^3 / \epsilon)^{1/2}$. For the vertical direction, the most accepted constant, $c = 0.2$, yields the present transition wavenumber being $\sim 0.06$ rad/m, around twice the smallest vertical wavenumber allowed in the model. We thus understand that the vertical spectrum is mostly dominated by the isotropic turbulence regime and the stratification effect is limited at a few smallest wavenumbers.

Even after the wave breaking occurs, the energy level significantly varies in a time scale longer than a typical wave period or an initial energy growth. A notable feature is that the energy in the horizontal mean flow component (Appendix C) gradually increases and finally dominates the whole system. This manifestation of vertically sheared horizontally large scale flow structure is a common character of stratified turbulence as identified in direct numerical simulations \citep{riley2003dynamics}.

\begin{figure}[t]
  \noindent\includegraphics[bb=0 0 624 741, width=0.8\textwidth]{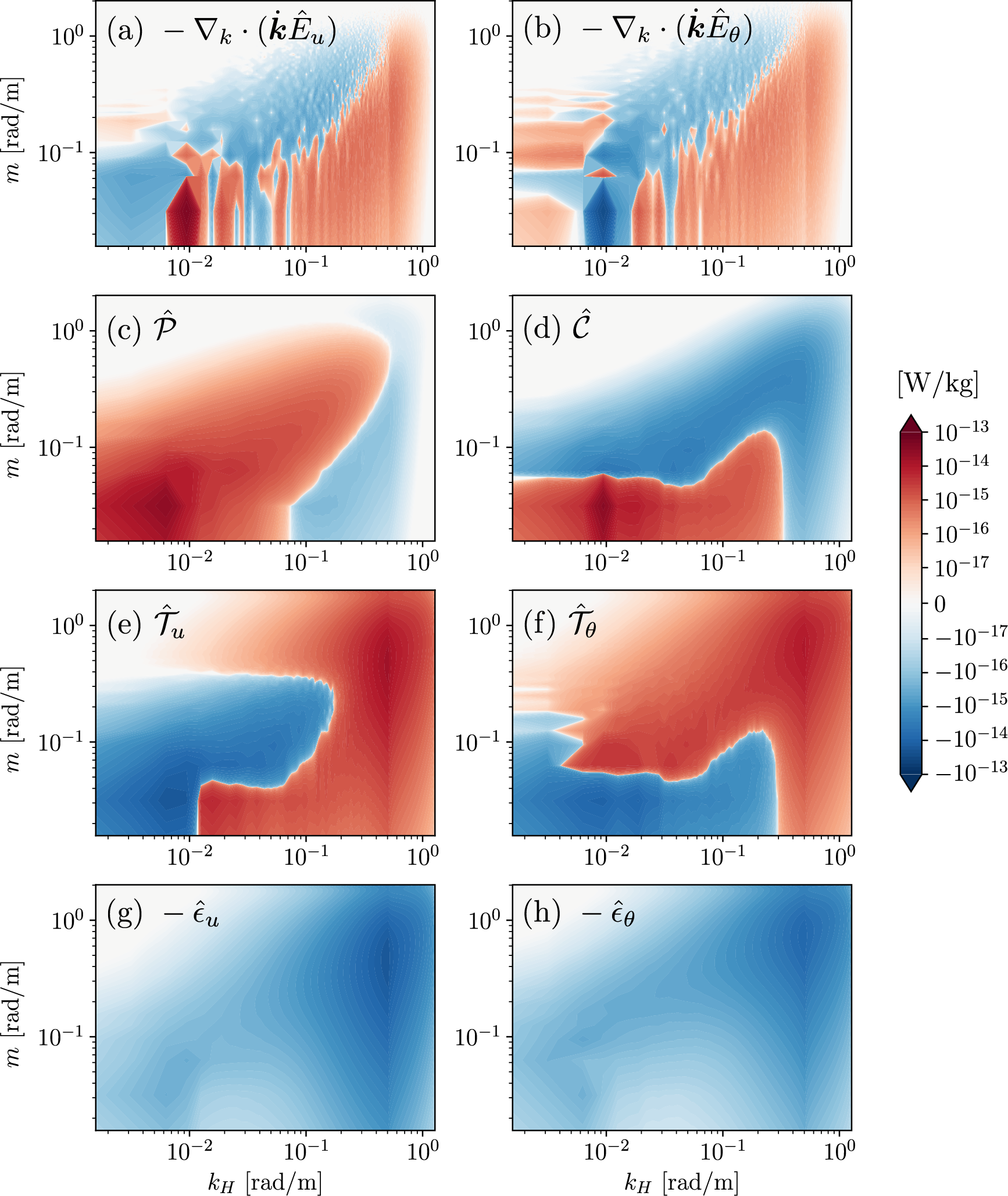}\\
  \caption{Terms in the energy equation in wavenumber space, (\ref{eq:budgets_wavenumber}), for the run \textbf{Id 2}. Each data is averaged over 8 days $\leq t <$ 10 days, integrated over the azimuthal angle, \yo{smoothed using a moving average for three points in the vertical direction, and multiplied by the wavenumbers $k_H$ and $m$ to create a content spectrum \citep{mccomas1981dynamic}.}} \label{fig:budget}
\end{figure}

To investigate the interscale energy transfer mechanisms in further detail, we evaluate the energy budgets in wavenumber space. For this analysis, we compute and store the terms in (\ref{eq:budgets_wavenumber}) during the simulation and integrate them for the azimuthal angle to create 2D plots against the horizontal wavenumber, $k_H$, and the vertical wavenumber, $m$ (Fig.~\ref{fig:budget}). The kinetic energy production rate, $\hat{\mathcal{P}}$, which originates from the lateral velocity shear in the background \yo{vortex}, is concentrated at the lowest wavenumber components that correspond to the primary mode (Fig.~\ref{fig:budget}c). Somewhat surprisingly, the energy production rate is negative in high horizontal wavenumbers, although it is negligibly small in the total energy budgets. Looking at the kinetic energy flux convergence term, $- \nabla_k \cdot (\dot{\boldsymbol{k}} \hat{E}_u)$, we find \yo{a} significant positive value at a horizontal wavenumber of $k_H \sim 10^{-2}$ rad/m that is slightly greater than the peak wavenumber in $\hat{\mathcal{P}}$, $k_H \sim 6 \times 10^{-3}$ rad/m. In light of the mechanism of elliptic instability, this energy transfer is interpreted to result from the stretching of horizontal wavenumber during the wave amplification process. This kinetic energy is concurrently converted to available potential energy via the vertical oscillation of the primary wave as evident in Fig.~\ref{fig:budget}d, in which $\hat{\mathcal{C}}$ is positive in the corresponding area.

We next examine the nonlinear interaction terms, $\hat{\mathcal{T}}_{\yo{u}}$ and $\hat{\mathcal{T}}_{\yo{\theta}}$, in Fig.~\ref{fig:budget}e, f. As expected from a basic picture of energy cascade, both $\hat{\mathcal{T}}_{\yo{u}}$ and $\hat{\mathcal{T}}_{\yo{\theta}}$ are negative at small horizontal and vertical wavenumbers and positive at large horizontal and vertical wavenumbers. However, the signs are different in the intermediate-wavenumber range. Specifically, $\hat{\mathcal{T}}_u$ is positive but $\hat{\mathcal{T}}_\theta$ is negative at the low-vertical-wavenumber region ($m \lesssim 0.05$ rad/m). This region corresponds to the area where $\hat{\mathcal{C}}$ is positive in Fig.~\ref{fig:budget}d. These results are explained in terms of the shear instability; in a high-vertical-shear and large-horizontal-scale flow, when the local gradient Richardson number falls below 0.25, horizontally undulating vertical motions are spontaneously enhanced through Kelvin-Helmholtz or Holmboe wave instability \citep{caulfield2021layering}. During this process, the kinetic energy in the large-scale motion is first supplied to the high-horizontal-wavenumber components (i.e., $\hat{\mathcal{T}}_u>0$) and subsequently converted to the available potential energy ($\hat{\mathcal{C}} > 0$). This available potential energy will be further transferred towards higher wavenumbers ($\hat{\mathcal{T}}_\theta < 0$). At the intermediate vertical wavenumbers ($m \gtrsim 0.05$ rad/m), in contrast, $\hat{\mathcal{T}}_u$ is negative but $\hat{\mathcal{T}}_\theta$ is positive, and the available potential energy is converted to the kinetic energy. These results are the features of the gravitational instability, in which a large-scale density structure is skewed to create a smaller-scale density overturn ($\hat{\mathcal{T}}_\theta > 0$) followed by the release of potential energy to vertical motions ($\hat{\mathcal{C}} < 0$) and the energy transfer to smaller-scale turbulence ($\hat{\mathcal{T}}_u < 0$). Passing through these two kinds of instability processes, energy cascades down towards the smallest scale, where significant dissipation of both the kinetic and available potential energy takes place (Fig.~\ref{fig:budget}g, h).

\subsection{Case II: weak instability}
\begin{figure}[t]
  \noindent\includegraphics[bb=0 0 406 485, width=0.5\textwidth]{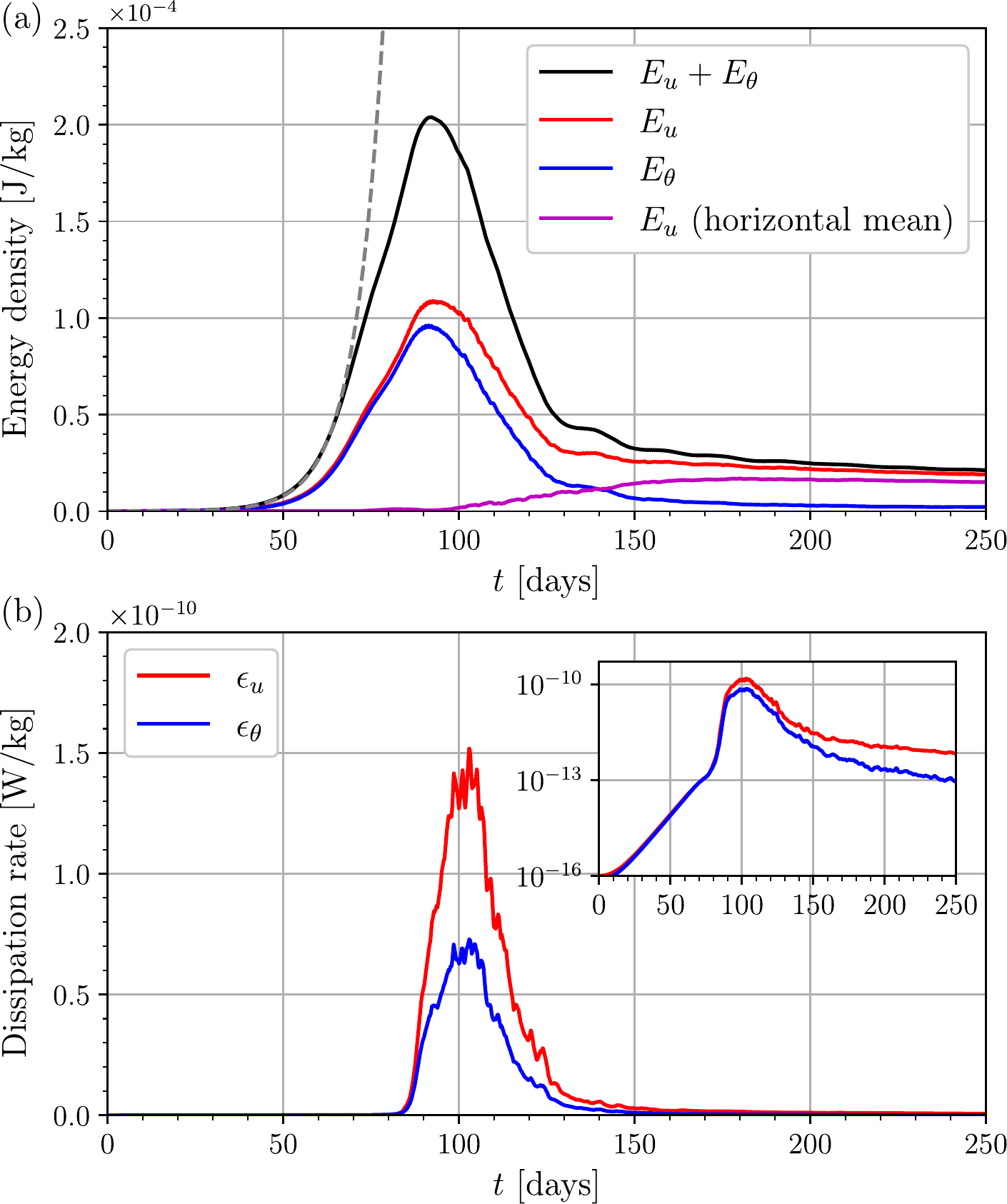}\\
  \caption{(a) Time series of the kinetic energy, available potential energy, their sum, and the kinetic energy of the horizontal mean ($k = \yo{\ell} = 0$) components for the run \textbf{Id 52}. Data are moving averaged over one period of vortex rotation. The grey broken line indicates the theoretically estimated energy growth rate. (b) Those of the kinetic energy and available potential energy dissipation rates. The inset shows the same data with a logarithmic scale in the vertical axis.} \label{fig:time_series_case_2}
\end{figure}

\begin{figure}[t]
  \noindent\includegraphics[bb=0 0 451 451, width=\textwidth]{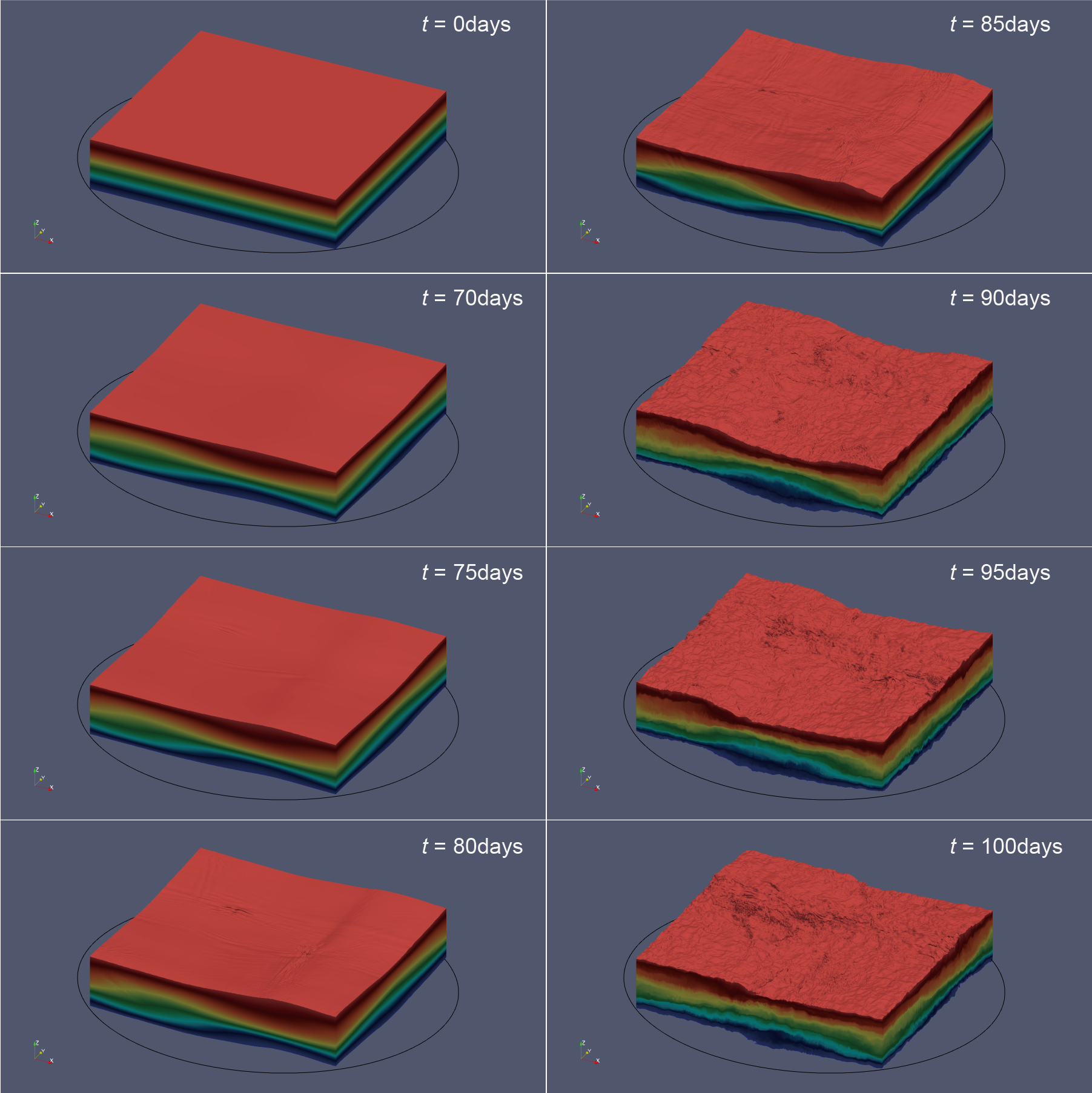}\\
  \caption{Evolution of the density field for the run \textbf{Id 52}. The vertical scale is magnified five times.} \label{fig:Rho_case_2}
\end{figure}

Next, we take up the case of $(N/f, Ro, e) = (10, 0.9, 0.1)$ (\textbf{Id 52} in Table 1), when the instability growth rate is one order smaller than Case I. A striking difference in the present results from the previous ones is the absence of clear signatures of the breaking of the primary wave. In the time series analysis (Fig.~\ref{fig:time_series_case_2}), after the initial stage of the exponential energy growth, the increasing tendency gradually weakens, and the energy smoothly transits to a decreasing phase. Figure~\ref{fig:Rho_case_2} illustrates the density structures before and after the peak of energy. In contrast to Fig.~\ref{fig:Rho_case_1}, density overturns are virtually invisible and, alternatively, small-scale wrinkles gradually appear on the isopycnal surface and they finally fill out the whole domain. As a result, the system continuously shifts from a laminar flow to a turbulent state. Even at the peak time of $\epsilon_u$, the typical size of the density overturn, namely the Thorpe scale, is several factors smaller than that of Case I. This point will be quantitatively discussed in Section 4.

\begin{figure}[t]
  \noindent\includegraphics[bb=0 0 720 430, width=\textwidth]{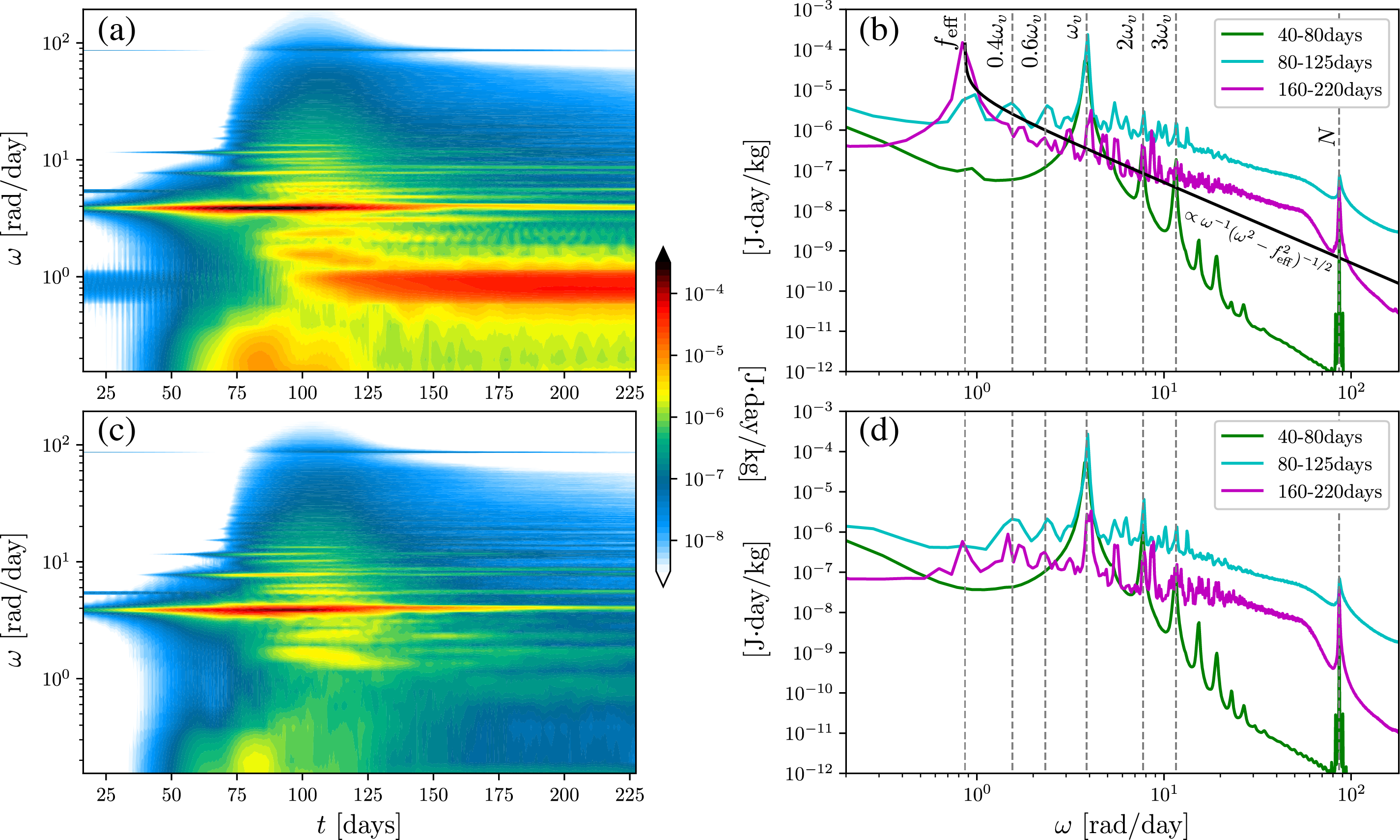}\\
  \caption{(a) Time-frequency diagram of the kinetic energy spectra for the run \textbf{Id 52}. Fourier transform is carried out for each time segment of 20 rotation periods ($\sim$32.5 days) at each location in the moving frame, $\boldsymbol{X}$, and the horizontal velocities are rotated as (\ref{eq:velocity_rotate}). (b) Energy spectra for certain intervals of time. Here, $f_{\rm eff} \equiv f (1 - Ro)$ represents the effective inertial frequency in the rotating frame. The black curve indicates the shape of the Garrett-Munk internal wave spectrum. (c, d) Those of the available potential energy spectra.} \label{fig:frequency_spectrum}
\end{figure}

To examine the temporal characteristics in further detail, we carry out time-frequency analysis \citep{brouzet2016energy}. Using a cosine-tapered window function, the time series of the kinetic energy and available potential energy spectra are computed as shown in Fig.~\ref{fig:frequency_spectrum}. At the initial stage ($t \lesssim  70$ days), most of the energy is contained at $\omega = \omega_v \sim 3.9$ rad/day, in agreement with the prediction from linear analysis in Section 2. After that, energy in the harmonics, $2\omega_v, 3\omega_v, \ldots$ gradually grows. Around $t = 90$ days, when the total energy is maximum, the situation abruptly changes; a pair of low-frequency components with $\omega \sim 0.4\omega_v, 0.6\omega_v$ rapidly gain energy. This is a signature of a parametric subharmonic instability of the primary $\omega_v$ component. Subsequently, after $t \sim 100$ days, \yo{the local inertial component, $\omega \sim f_{\rm eff} \equiv f (1 - Ro)$,} corresponding to the pure horizontal flow (Appendix C) grows and the energy in the other components steadily declines. \yo{Conducting bispectrum analysis \citep{yang2022energy}, we have confirmed that a net energy transfer from a high-frequency range ($\lesssim$ 10 rad/day) to the inertial component takes place around 80 days $< t <$ 160 days (results not shown). Drastic changes in spectral shapes} are also evident in Fig.~\ref{fig:frequency_spectrum}c,~d, which demonstrate the frequency spectra for three phases in the simulation. At the end of the simulation, most of the energy is concentrated at $\omega \sim f_{\rm eff}$ in the kinetic energy part, and the spectral form is close to the Garrett-Munk spectrum, a typical internal wave spectrum in the ocean \citep{garrett1972space,garrett1975space}, although a slight difference in the spectral slope is confirmed at the high-frequency region.

Succession of resonances to achieve wave turbulence found here is analogous to those reported in various contexts including internal wave attractor experiments \citep{brouzet2016energy,brouzet2016internal,davis2020succession} and numerical or tank experiments of elliptic instability for internal or inertial waves \citep{le2017inertial,le2018parametric,le2019experimental}. However, in the present study, the final state of the spectrum differs from these previous ones. In the pure internal wave case, energy was distributed broadly in frequency space. In pure rotation cases, the final states were either inertial wave turbulence or a strong geostrophic vortex. Presumably, existence of lateral boundary or absence of stratification prevented the excitation of a horizontally homogeneous flow mode. Energy concentration to the vertically sheared horizontal flow is a unique feature in the present \yo{setup} relevant to ocean conditions.

Overall, in Case II, energy transfer in spectral space is dominated by weakly nonlinear processes that involve high-harmonic generation and subharmonic instabilities. Wave breaking that causes high energy dissipation takes place at spatial scales much smaller than the primary unstable mode. This result is in contrast to Case I, in which the primary wave directly breaks through gravitational instability; the system is totally governed by strongly nonlinear processes. \yo{Comparing the two results of time series, Fig.~\ref{fig:time_series_case_1} and Fig.~\ref{fig:time_series_case_2}, we notice that the maximum levels of energy density and dissipation rates are greater for Case II contrary to the smaller growth rate. This seemingly surprising result is ascribed to the difference of the background buoyancy frequency; $N/f=3$ for Case I and $N/f=10$ for Case II. This point will be further investigated in the following section.}

\section{Discussion on parameter dependence of turbulence properties}
Results in Section 3 have shown that the energy saturation processes of internal waves excited by AAI can be qualitatively classified into two categories; direct wave breaking v.s. weakly nonlinear wave-wave interactions. In this section, we aim to distinguish these two scenarios quantitatively using a suit of simulation results, total 56 runs from \textbf{Id 3} to \textbf{Id 58}. A main direction is to parametrize the kinetic energy dissipation rate, $\epsilon_u$, in terms of parameters precisely determined from the external conditions, such as the linear instability growth rate, $\lambda$ (abbreviated from $\lambda^\nu_{\rm max}$), and rationalize it on a theoretical basis. Here, since the state of the system varies significantly during a long-term simulation, to provide a discussion on a specific basis, we set the final time of the simulation as $t_{\rm fin} = 15 / \lambda$ (this is enough to cover the stage of wave breaking and transition to turbulent states for all the experiments), \yo{took} a running mean of scalar data such as $\epsilon_u$ over one rotation period $T$, and used the peak value of it for the analysis.
\clearpage
\begin{figure}[t]
  \noindent\includegraphics[bb=0 0 371 762, width=0.5\textwidth]{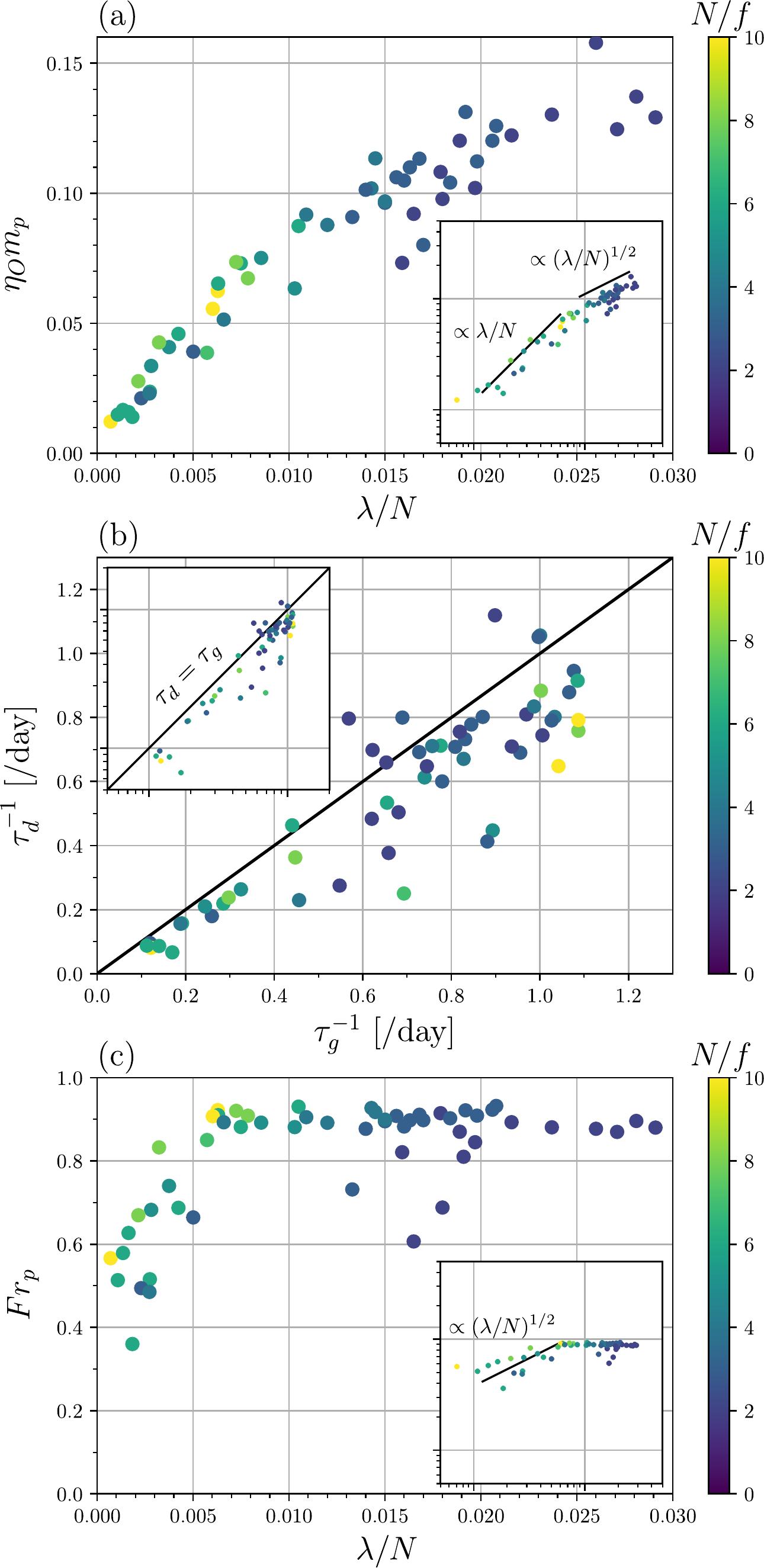}\\
  \caption{(a) Scatter plot of the Ozmidov length scale multiplied by the vertical wavenumber of the primary wave, $\eta_O m_p$, v.s. the instability growth rate devided by the buoyancy frequency, $\lambda / N$. (b) Scatter plot of the inverses of the primary wave energy growth and decay time scales. (c) Scatter plot of the Froude number defined for the primary unstable wave, $Fr_p$, v.s. $\lambda / N$. The insets are the log-log plots. The colors represent the buoyancy frequency divided by the Coriolis parameter, $N/f$.} \label{fig:scatter_lambda}
\end{figure}
\clearpage
We are aiming at finding a scaling relationship between an external parameter and the energy dissipation rates. For this purpose, a dimensionless quantity is more favorable than the original dimensional one. Specifically, we define the Ozmidov length scale multiplied by the vertical wavenumber of the primary wave as
\begin{align}
\eta_O m_p \equiv \left( \frac{\epsilon_u}{N^3} \right)^{1/2} \left( \frac{2 \pi}{L_z} \right)
\end{align}
and utilize it instead of $\epsilon_u$. First, Fig.~\ref{fig:scatter_lambda}a demonstrates the scatter plot between $\lambda / N$ and $\eta_O m_p$. The significant positive correlation is naturally understood that stronger instability results in higher energy dissipation. Besides, the log-log plot indicates two types of scaling relationships,
\begin{align} \label{eq:scale1}
\begin{cases}
\eta_O m_p \propto \lambda / N \quad & \text{i.e.,} \quad \epsilon_u \propto L_z^2 N \lambda^2 \quad \text{for} \quad \lambda / N \lesssim 0.008 \\
\eta_O m_p \propto (\lambda / N)^{1/2} \quad & \text{i.e.,} \quad \epsilon_u \propto L_z^2 N^2 \lambda \quad \text{for} \quad \lambda / N \gtrsim 0.008 .
\end{cases}
\end{align}
We shall call the two regimes as ``weak instability'' and ``strong instability'' and try to provide a reasoning of (\ref{eq:scale1}) in the following analysis.

Now, we focus on the energy density in the primary mode, denoted as $E_p$. For the evolution of $E_p$, we may define two types of time scales; the energy growth time scale, $\tau_g$, and the energy decay time scale, $\tau_d$. The growth time scale naturally derives from the linear instability growth rate as $\tau_g = 1 / (2 \lambda)$, where the factor two is placed because the energy is proportional to the square of the wave amplitude. The decay time scale is rather difficult to define precisely because it involves complicated nonlinear processes. Here, making an assumption that the energy lost from the primary wave will be dissipated without escaping outside the system, we write the decay time scale as $\tau_d = E_p / \epsilon_u$. Figure~\ref{fig:scatter_lambda}b is a scatter plot against $1/\tau_g$ and $1/\tau_d$. The accumulation of data along the diagonal line with the correlation coefficient of \yo{0.884} supports the hypothesis that the growth and decay time scales, $\tau_g$ and $\tau_d$, almost coincide for all the experimental conditions.

Using the primary mode energy density, $E_p$, we further introduce a dimensionless parameter,
\begin{align}
Fr_p = \frac{m_p (2 E_p)^{1/2}}{N} ,
\end{align}
which represents the Froude number defined for the primary mode. As shown in Fig.~\ref{fig:scatter_lambda}c, in every simulation, $Fr_p$ is strictly bounded around 0.9. This value may represent the condition for the direct wave breaking of the primary mode. When $\lambda / N$ is small, $Fr_p$ does not reach such an upper bound, presumably because the weakly nonlinear interactions take place to drain energy from the primary mode before its amplitude grows enough. In the log-log plot, although the deviation is large, the most plausible functional relationship appears to be $Fr_p \propto (\lambda / N)^{1/2}$ for the weak-instability regime. Then, accepting the two scaling relationships,
\begin{align}
\begin{cases}
Fr_p \propto \left( \lambda / N \right)^{1/2} \quad & \text{for} \quad \lambda / N \lesssim 0.008 \\
Fr_p \sim \text{constant} \quad & \text{for} \quad \lambda / N \gtrsim 0.008 ,
\end{cases}
\end{align}
and using $\tau_d \sim \tau_g$, we derive the expected scaling laws, (\ref{eq:scale1}).

Why $Fr_p \propto \left( \lambda / N \right)^{1/2}$ holds for the weak-instability regime? This can be explained from a basic knowledge of weak turbulence theory \citep{nazarenko2011wave}. In general, in a weakly nonlinear wave system where energy is transferred in spectral space through resonant interactions, the typical time scale that characterizes the evolution of wave energy density is called the kinetic time scale, $\tau_k$. For internal wave turbulence, resonant triad interactions dominate the energy transfer in spectral space, and the net energy transfer rate divided by the energy density is quadratic of the wave amplitude \citep{lvov2012resonant,onuki2019parametric}. In the present case, assuming that the typical wave amplitude is scaled by $Fr_p$ and the typical wave frequency is scaled by $N$, we can write the kinetic time scale as $\tau_k \sim Fr_p^{-2} / N$. As is clear from its definition, the energy decay time scale of the primary mode is equivalent to the kinetic time scale; i.e, $\tau_d \sim \tau_k$ establishes. Finally, again using $\tau_g \sim \tau_d$, we arrive at the aforementioned scaling relationship, $Fr_p \propto (\lambda / N)^{1/2}$.

Our classification of the scaling relationships has several common aspects with the wave-turbulence transition model (WT model, hereafter) proposed by \cite{d2000wave}. Indeed, the \yo{dependence} of $\epsilon_u$ on wave amplitudes for the weak instability regime well corresponds to the low-wave-energy regime in the WT model. This scaling is also consistent with the fine-scale parameterization widely used to infer energy dissipation rates in the open ocean \citep{gregg1989scaling,polzin1995finescale,ijichi2015frequency}. However, the present scaling of the strong instability regime differs from the WT model. In our simulations, energy in the largest-scale waves is limited by the condition of $Fr_p \sim 1$, and $\epsilon_u$ is controlled by the instability growth rate that determines the characteristic time scale. For a high energy regime of the WT model, on the other hand, the time scale is prescribed by the Coriolis frequency and $\epsilon_u$ is proportional to a variable wave energy level. This proportionality between the kinetic energy and the energy dissipation rate is an accepted notion underlying the large-eddy method to infer $\epsilon_u$ from velocity measurement \citep{moum1996energy,beaird2012dissipation}. On this point, our scaling is a new one that may apply to special situations where internal waves are generated from balanced motion through ageostrophic instability.

\begin{figure}[t]
  \noindent\includegraphics[bb=0 0 380 290, width=0.5\textwidth]{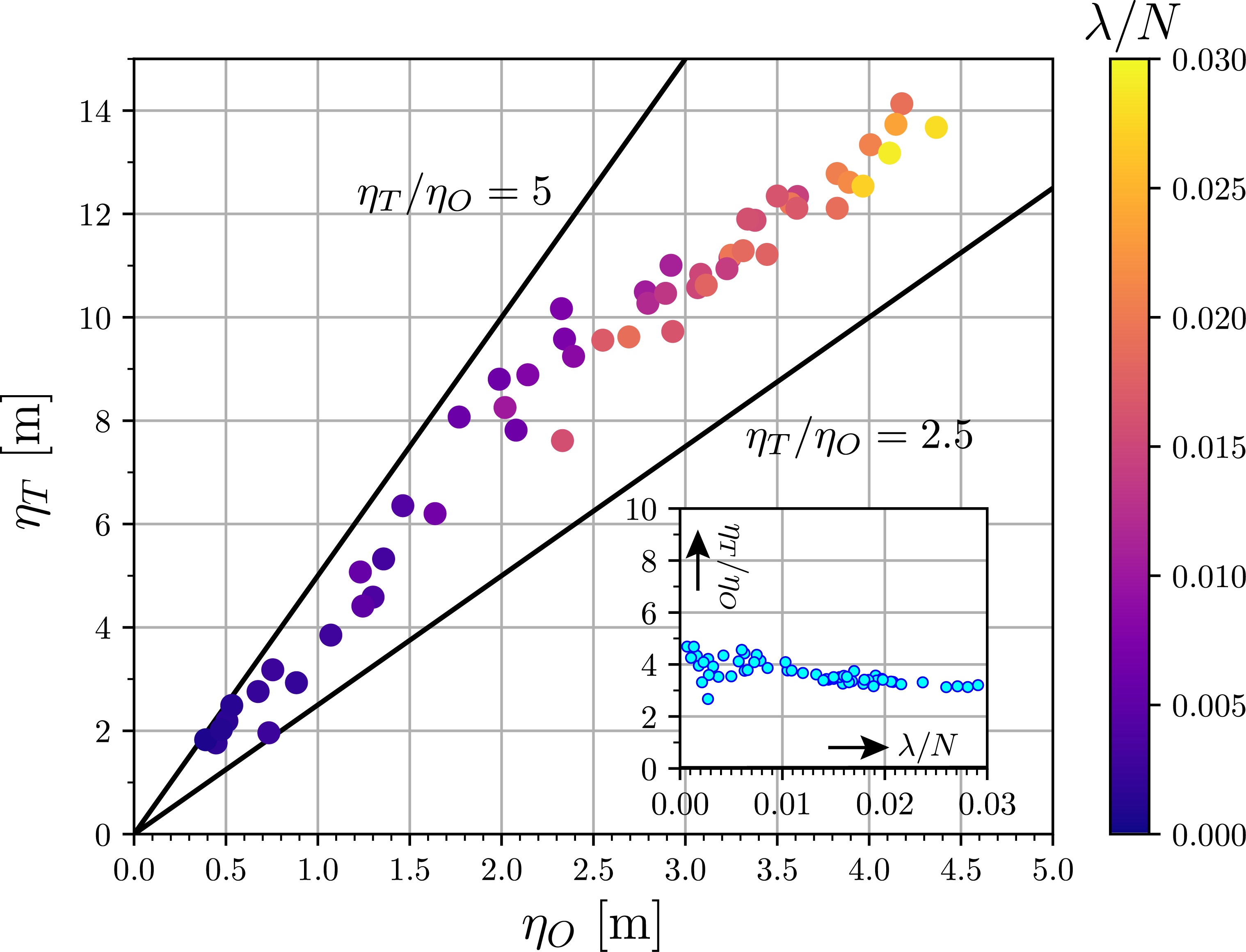}\\
  \caption{Scatter plot of the Ozmidov length scale, $\eta_O$, v.s. the Thorpe length scale, $\eta_T$. The color represents the instability growth rate divided by the buoyancy frequency, $\lambda / N$. The inset is the scatter plot of $\lambda / N$ v.s. $\eta_T / \eta_O$.} \label{fig:Ozmidov_Thorpe}
\end{figure}

In the context of ocean energetics, the ratio between the kinetic energy dissipation rate and the available potential energy dissipation rate, i.e., the mixing coefficient $\Gamma \equiv \epsilon_\theta / \epsilon_u$, is of special importance \citep{de2016impact,gregg2018mixing}. However, to obtain an accurate estimate of $\Gamma$ in a numerical model, the number of \yo{grid points} should be high enough to resolve the smallest scales of \yo{turbulent} motion where the molecular viscosity and diffusivity dominate. In the present model, since a subgrid scheme is employed to represent eddy viscosity and diffusivity, the mixing coefficient is not directly available. Alternatively, we investigate the Thorpe scale, $\eta_T$, the typical size of density overturn calculated by sorting the instantaneous vertical density profiles. Scaling estimates of stratified turbulence argue that $\Gamma$ is a function of the ratio between the two length scales, $\eta_O$ and $\eta_T$ \citep{garanaik2019inference}. In the ocean, observation data revealed that the value of $\eta_T / \eta_O$ ranges over two orders of magnitude and $\Gamma$ varies accordingly as $\Gamma \propto (\eta_T / \eta_O)^{4/3}$ \citep{ijichi2018observed,ijichi2020variable}. In \yo{our} experiments, although $\eta_O$ and $\eta_T$ both vary depending on $\lambda/N$ over one decade, the variation of their ratio is much smaller, around 2.5-5 throughout all the simulations (Fig.~\ref{fig:Ozmidov_Thorpe}). This result implies that the mixing coefficient is irrelevant to the wave breaking scenarios in the present \yo{situation}.

\begin{figure}[t]
  \noindent\includegraphics[bb=0 0 418 811, width=0.5\textwidth]{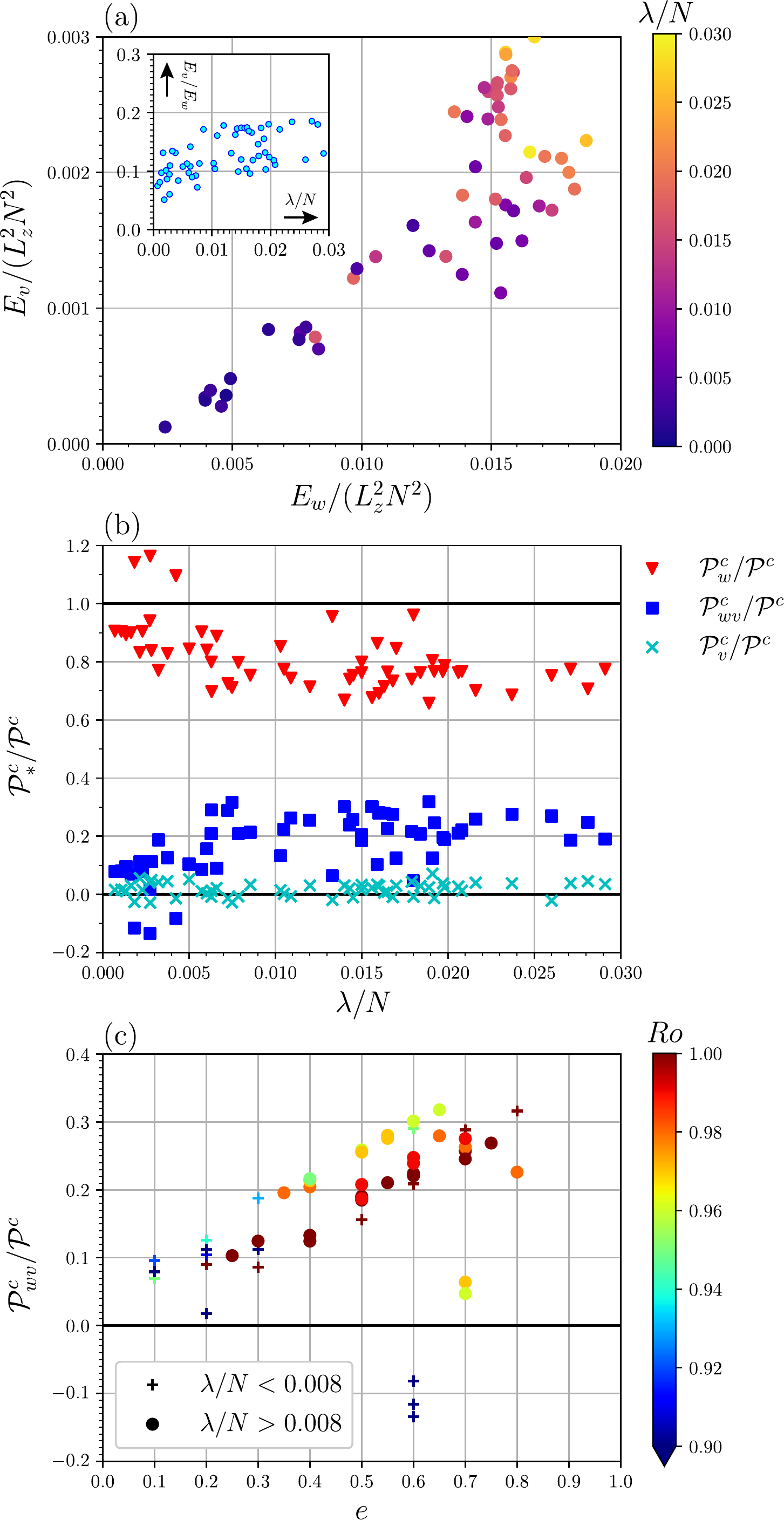}\\
  \caption{(a) Scatter plot of the wave mode energy, $E_w$, v.s. the vortical mode energy, $E_v$. Both are scaled using the vertical length of the domain, $L_z$, and the buoyancy frequency, $N$. The color represents the instability growth rate, $\lambda$, divided by $N$. The inset is the scatter plot of $\lambda / N$ v.s. $E_v / E_w$. (b) The cumulative contributions to the kinetic energy production from the terms defined in (\ref{eq:production_separate}) are plotted against $\lambda/N$. (c) The contribution from $\mathcal{P}^c_{wv}$ is plotted against the ellipticity of the background vortex, $e$. Different symbols are used to distinguish the weak and strong instability regimes. The color represents the Rossby number. In (b) and (c), $\mathcal{P}^c \equiv \mathcal{P}_{ww}^c + \mathcal{P}_{wv}^c + \mathcal{P}_{v}^c$ represents the total amount of kinetic energy production.} \label{fig:wave_vortex}
\end{figure}

At the end of discussion, we analyze the role of the vortical mode in the energy budget. Following the formulation of Section 2, we separate $(\boldsymbol{u}, \theta)$ data into the wave part and the vortical part and compute the energy density for each part, $E_v$ and $E_w$, respectively. The energy production terms, $\mathcal{P}_{ww}, \mathcal{P}_{wv}$, and $\mathcal{P}_v$, are also computed and integrated over time as
\begin{align}
(\mathcal{P}^c_{ww}, \mathcal{P}^c_{wv}, \mathcal{P}^c_v) = \int_0^{t_{\rm fin}} (\mathcal{P}_{ww}, \mathcal{P}_{wv}, \mathcal{P}_v) dt
\end{align}
to discuss their cumulative contributions. As shown in Fig.~\ref{fig:wave_vortex}a, the peak value of the vortical mode energy is no more than 20\% of the wave mode energy. Although a positive correlation between $\lambda / N$ and $E_v / E_w$ may exist, the deviation is too large to derive a specific scaling relationship. Among the energy production terms, as has been expected in linear analysis, the dominant contribution comes from $\mathcal{P}^c_{ww}$ (Fig.~\ref{fig:wave_vortex}b). For the remaining two terms, while $\mathcal{P}^c_v$ is ignorable in every case, $\mathcal{P}^c_{wv}$ explains up to 30\% of the total energy production rate. Therefore, although the vortical mode does not gain energy directly from the reference flow, it effectively assists the energy production of the wave mode components. A unique character of $\mathcal{P}^c_{wv}$ is that it depends on the Rossby number $Ro$ and the ellipticity $e$ of the vortex rather than $\lambda / N$. As revealed in Fig.~\ref{fig:wave_vortex}c, $\mathcal{P}^c_{wv} / \mathcal{P}^c$ tends to increase with $e$. This functional relationship is particularly clear when $Ro$ is close to 1. As $Ro$ departs from 1, this tendency becomes obscure, and $\mathcal{P}^c_{wv}$ can even be negative for $e=0.6$ and $Ro \lesssim 0.9$ cases. We have thus assert that the IGWs generation assisted from the vortical mode is effective when both the Rossby number and the ellipticity are large, which is consistent with the theoretical analysis of \cite{mcwilliams1998fluctuation}. When $Ro$ is small, the vortical mode disturbances can have both the positive and negative impacts on the wave energy production. A proper explanation of this unexpected result is yet to be given at the present stage.
\clearpage
\clearpage
\section{Conclusions}
Spontaneous generation of internal gravity waves from a balanced motion is an important process for the ocean and atmosphere dynamics. In this study, we have theoretically and numerically investigated a kind of wave generation phenomenon named ageostrophic anticyclonic instability (AAI) in a specific situation of spatially uniform elliptic vortex. Using a time-dependent Fourier transform method, we have clarified temporal characteristics and energetic properties of growing internal waves in this unique setting.

Performing a series of numerical simulations, we have identified two types of wave breaking scenarios and categorized them using the linear instability growth rate scaled by the background buoyancy frequency, $\lambda / N$. When $\lambda / N \gtrsim 0.008$, the Froude number, $Fr_p$, defined for the primary unstable mode resulting from AAI promptly reaches around unity, thus directly \yo{leading} to gravitational instability of the primary wave. When $\lambda / N \lesssim 0.008$, in contrast, before $Fr_p$ grows sufficiently large, weakly nonlinear wave-wave interactions arise to redistribute energy across the frequency spectrum. Consequently, the system gets into a weak turbulence state and the wave breaking occurs at vertical scales much smaller than the primary unstable wave.

Throughout the experimental conditions, the growth and decay time scales of the primary wave energy are always comparable. Besides, in the weak turbulence regime, this time scale is roughly proportional to $Fr^{-2}_p$, while in the strong instability regime $Fr_p$ is almost constant and irrelevant to any time scales. Combining these two estimates, we derive distinct scaling relationships between the linear instability growth rate $\lambda$ and the energy dissipation rate $\epsilon_u$, for the strong- and weak-instability regimes, respectively. Notably, in the strong instability regime, $\epsilon_u$ does not depend on the internal wave energy level. This result challenges the conventional parameterization schemes in which $\epsilon_u$ is parameterized as a function of a velocity spectral level at the internal wave band.

\yo{This study has considered a particular type of AAI; parametric excitation of IGWs from a fully balanced homogeneous flow of an elliptic shape, in a highly idealized setting. In the real ocean, a submesoscale eddy may not maintain a coherent structure over such a long period as $O(10)$days, and there should be velocity gradients in both the horizontal and vertical directions. Recently, \cite{eden2019gravity} and \cite{chouksey2022gravity} carried out numerical simulations on gravity wave emission from an unstable baroclinic shear flow. In their experiments, however, the generated wave energy was quite weak because the Rossby number was small ($Ro \leq 0.3$), and the wave dissipation processes were not fully investigated. Gravity wave generation forced by evolving vortical motion is another candidate for the energy sink of the balanced mode \citep{sugimoto2015cyclone,sugimoto2016generation}. It would be worth assessing the scaling for energy dissipation rates also in these processes.

Finally, for the actual ocean situations, it is important to consider more intense types of instability\textemdash symmetric/centrifugal instabilities occurring when the potential vorticity (PV) changes sign. In recent years, a series of field measurements in the Kuroshio region have revealed that the interplay of the intense instabilities in negative PV and internal waves trapped within positive but low PV creates a wide-ranging turbulent mixing zone off the southern coast of Kyushu Island, Japan \citep{tsutsumi2017turbulent,nagai2017first,nagai2021kuroshio}. Extending our understanding on energetics of ageostrophic instability from a weak internal wave-emission regime to a much stronger instability regime remains a challenging problem for a future study.}

\clearpage
\acknowledgments
We are grateful to Anne Takahashi for the discussion on scaling relationships of stratified turbulence. \yo{We also appreciate helpful comments from two anonymous reviewers.} This study was supported by JSPS Overseas Research Fellowship as well as KAKENHI Grant Number JP20K14556, and in part by Collaborative Research Program of the Research Institute for Applied Mechanics, Kyushu University. For the numerical calculations, we used the FUJITSU Supercomputer PRIMEHPC FX1000 and FUJITSU Server PRIMERGY GX2570 (Wisteria/BDEC-01) at the Information Technology Center, The University of Tokyo. We gratefully acknowledge support from the PSMN (Pôle Scientifique de Modélisation Numérique) of the ENS de Lyon for the computing resources.

%
%
\datastatement
The numerical simulation codes are originally developed by the first author, Yohei Onuki. The experimental results are too large to be distributed, but the basic Fortran codes, parameter input files, and Python scripts for postprocessing are accessible at https://github.com/yonuki-models/numerical-simulation-aai.









%



\appendix[A]
\appendixtitle{Subgrid models of the viscosity and diffusivity}
In the nonlinear simulations, we employ scale-dependent eddy viscosity and diffusivity as subgrid models to keep the virtual Reynolds number at the low-wavenumber region sufficiently high while avoiding excessive energy accumulation at the high-wavenumber regions. Emulating the formulation of spectral eddy viscosity of classical large-eddy simulations \citep{chollet1981parameterization,domaradzki2021large}, the viscosity coefficients are varied depending on the spectral energy density at the truncation wavenumbers. Since the grid spacing is highly anisotropic in the present case, we modify the formulation originally invented for isotropic turbulence, borrowing the technique of \cite{furue2003energy}. We first define scaled horizontal and vertical wavenumbers, $k_s \equiv (k^2/\tilde{k}^2_c + \ell^2/\tilde{\ell}^2_c)^{0.5}$ and $m_s \equiv |m / \tilde{m}_c|$. Then, eddy viscosity and diffusivity are arranged as
\begin{align}
\nu_{ue} = \nu_{\rho e} = C_H \sqrt{\frac{E_{\yo{u}}(k_s = 1)}{\tilde{k}_c}} k^2_s + C_V \sqrt{\frac{E_{\yo{u}}(m_s = 1)}{\tilde{m}_c}} m_s^4 ,
\end{align}
where $E_{\yo{u}}(k_s = 1) \equiv \sum_{0.9 < k_s \leq 1} \hat{E}_{\yo{u}} (\boldsymbol{k}) / 0.1 \tilde{k}_c$ and $E_{\yo{u}} (m_s = 1) \equiv \sum_{0.9 < m_s \leq 1} \hat{E}_{\yo{u}} (\boldsymbol{k}) / 0.1 \tilde{m}_c$ are the 1d spectral density \yo{of kinetic energy} averaged over thin shells of the side and a pair of vertical faces of an elliptic cylinder composed of the effective wavenumbers, respectively. This parameterization involves dimensionless adjustable parameters, $C_H$ and $C_V$.

For the calibration purpose, we have performed a high-resolution experiment without using the subgrid model, i.e., direct numerical simulation (DNS) and also a relatively low-resolution experiment using the subgrid model, i.e., large eddy simulation (LES), in an equivalent setting, which are the runs \textbf{Id 1, 2} in Table 1. It was found that the results of DNS and LES well coincide (Fig.~A1) by setting the dimensionless constants as $C_H = 1$ and $C_V = 1.5$. Then, we adopt these values throughout the other simulations.

\begin{figure}[t]
  \noindent\includegraphics[bb=0 0 415 491, width=0.5\textwidth]{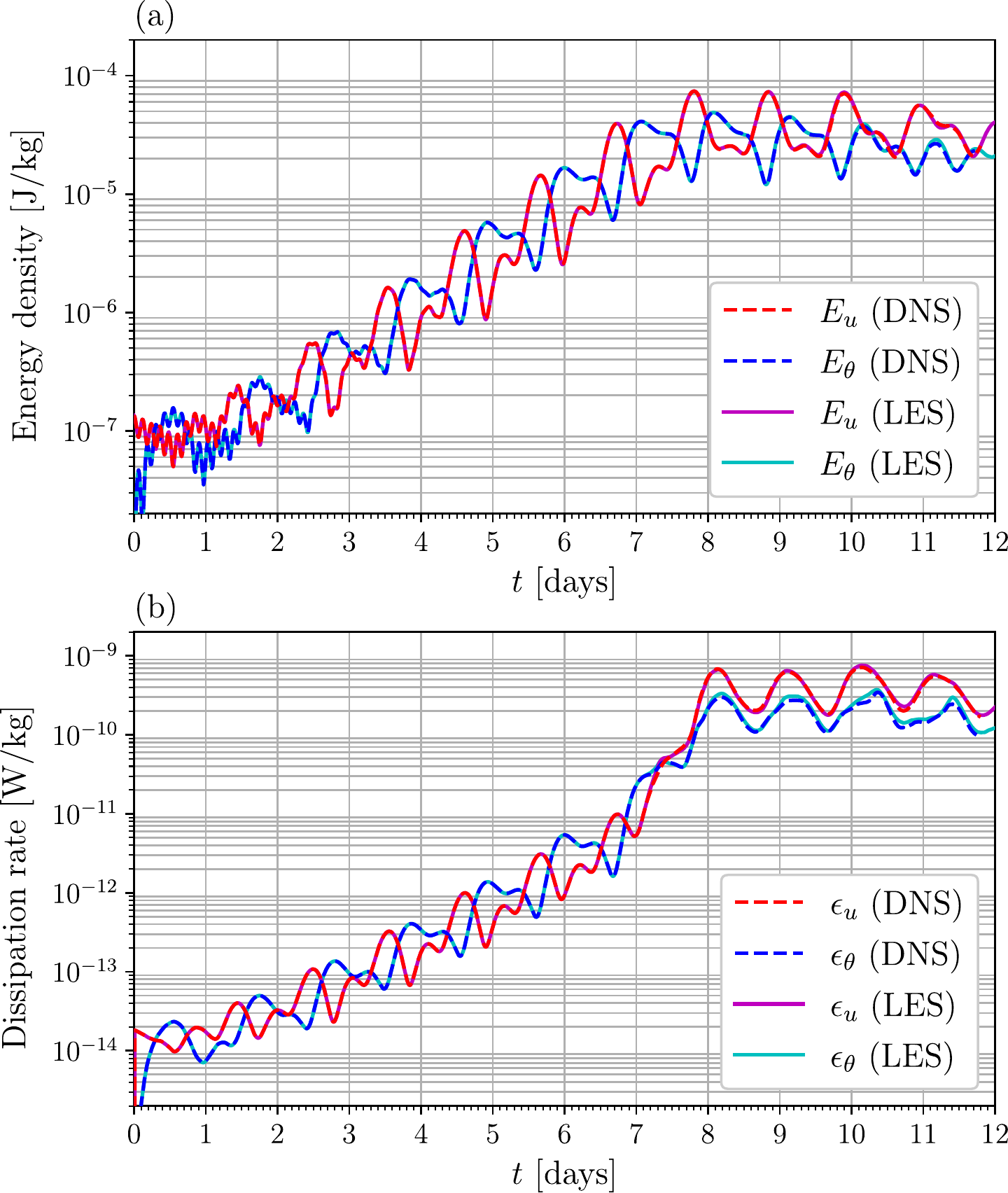}\\
  \caption{(a) Time series of the kinetic energy, available potential energy, and their sum for the runs \textbf{Id 1} (DNS) and \textbf{Id 2} (LES). In \textbf{Id 1}, the number of the grid points is $3072 \times 3072 \times 384$ and the viscosity and diffusivity are fixed. In \textbf{Id 2}, the number of the grid points is $768 \times 768 \times 192$ and the subgrid model for the viscosity and diffusivity is implemented. (b) Those of the kinetic energy and available potential energy dissipation rates. Note that the vertical axis was in a linear scale for Fig. \ref{fig:time_series_case_1} while it is in a logarithmic scale in the present plots.} \label{fig:DNS_compare}
\end{figure}

\appendix[B]
\appendixtitle{\yo{Dependence} of linear stability analysis on $N/f$}
In the present study, we have employed the stratification parameter as $N/f=10$ for the linear analysis while varied it over $2 \leq N/f \leq 10$ for nonlinear simulations. In light of the actual oceanic conditions, these values are relatively small. At the thermocline of low-latitude oceans, $N/f$ occasionally reaches around 100. This discrepancy of $N/f$ is not a serious problem for linear stability analysis because a scaling relationship exists between $N/f$ and $\tan \phi$ \citep{aspden2009elliptical}. Figure B1(a) demonstrates the instability growth rate, $\lambda$, against $N/f$ and $\tan \phi$. Clearly, the unstable regions follow lines on which $(N/f)/\tan \phi$ is a constant. Figure~B1(b) shows $\lambda_{\rm max}$ as a function of $N/f$. We confirm that the maximum growth rates hardly depend on $N/f$ when $N/f \geq 2$. To summarize, a choice of $N/f = 10$ is enough to check the parameter \yo{dependence} of linear stability analysis as long as $\tan \phi$ is varied over a wide range.

\begin{figure}[t]
  \noindent\includegraphics[bb=0 0 426 491, width=0.5\textwidth]{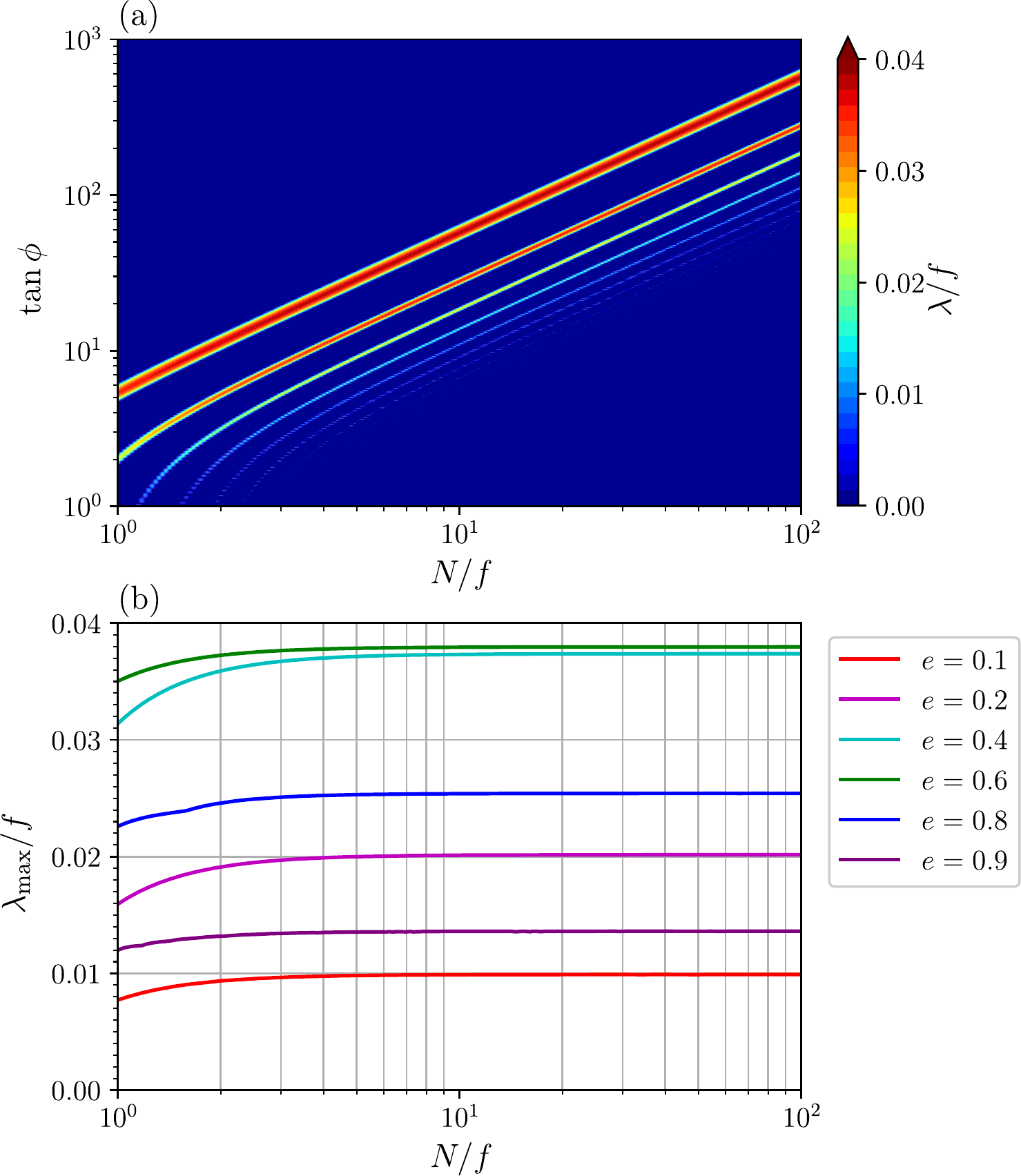}\\
  \caption{(a) Numerically obtained instability growth rates, $\lambda$, scaled by $f$ for $Ro = 0.95$ and $e = 0.6$ are contoured against $N/f$ and $\tan \phi$. (b) The maximum instability growth rates in wavenumber space, $\lambda_{\rm max}$, scaled by $f$ for $Ro = 0.95$ and several $e$ are plotted against $N/f$.} \label{fig:floquet_against_N}
\end{figure}

\appendix[C]
\appendixtitle{Pure horizontal flow mode}
Equation (\ref{eq:linear}) allows a solution of the vertically sheared horizontal flow mode to be computed analytically. Setting $k = \ell = 0$ and $\hat{w} = 0$, we may write the equations as
\begin{subequations}
\begin{align}
\frac{d \hat{u}}{dt} & = (f - \beta) \hat{v} \\
\frac{d \hat{v}}{dt} & = (\alpha - f) \hat{u} .
\end{align}
\end{subequations}
Since $0 < \alpha, \beta < f$, these equations are immediately solved as
\begin{subequations}
\begin{align}
\hat{u} & = \frac{A \cos (\omega_h t + \vartheta)}{(1 - \alpha / f)^{1/2}} \\
\hat{v} & = - \frac{A \sin (\omega_h t + \vartheta)}{(1 - \beta / f)^{1/2}} ,
\end{align}
\end{subequations}
where $\omega_h \equiv f (1 - \alpha / f)^{1/2} (1 - \beta / f)^{1/2}$ is the natural frequency of this mode in the frame fixed to the \yo{Earth}, and $A$ and $\vartheta$ are arbitrary constants. For this mode, the buoyancy $\hat{\theta}$ is arbitrary chosen.

In a frame rotating with the vortex, the horizontal velocities are changed according to (\ref{eq:velocity_rotate}) as
\begin{subequations}
\begin{align}
\hat{u}^r & \sim \frac{A \left\{ \cos (\omega_h t + \omega_v t + \vartheta) + \cos (\omega_h t - \omega_v t + \vartheta) \right\}}{2 (1 - \alpha / f)^{1/2}} - \frac{A \left\{ \cos (\omega_h t + \omega_v t + \vartheta) - \cos (\omega_h t - \omega_v t + \vartheta) \right\}}{2 (1 - \beta / f)^{1/2}} \\
\hat{v}^r & \sim \frac{A \left\{ \sin (\omega_h t + \omega_v t + \vartheta) - \sin (\omega_h t - \omega_v t + \vartheta) \right\}}{2 (1 - \alpha / f)^{1/2}} - \frac{A \left\{ \sin (\omega_h t + \omega_v t + \vartheta) + \sin (\omega_h t - \omega_v t + \vartheta) \right\}}{2 (1 - \beta / f)^{1/2}} .
\end{align}
\end{subequations}
If the shape of the vortex is close to circular, or equivalently $e \ll 1$, while $Ro \sim O(1)$, we may expand the parameters as $\alpha / f = Ro / 2 + O(e)$ and $\beta / f = Ro / 2 + O(e)$. Consequently, we obtain
\begin{subequations}
\begin{align}
\hat{u}^r & = \frac{A \cos (f(1 - Ro) t + \vartheta) }{(1 - Ro/2)^{1/2}} + O(e) \\
\hat{v}^r & = - \frac{A \sin (f(1 - Ro) t + \vartheta) }{(1 - Ro/2)^{1/2}} + O(e)
\end{align}
\end{subequations}
and thus understand that $f_{\rm eff} \equiv f (1 - Ro)$ represents the effective inertial frequency that is the lower bound of the internal wave frequency in this \yo{setup}.


\bibliographystyle{ametsocV6}
\bibliography{references}

\end{document}